\documentclass{aa}
\usepackage{natbib}
\usepackage[binary-units=true]{siunitx}
\usepackage{subcaption}
\bibpunct{(}{)}{;}{a}{}{,}

\usepackage{graphicx}
\usepackage{txfonts}

\newcommand{\hrieuv}{HRI\textsubscript{EUV}\xspace}
\newcommand{\hrilya}{HRI\textsubscript{Lya}\xspace}

\DeclareMathAlphabet{\mathbfl}{OML}{cmm}{b}{it} 
\newcommand{\vect}[1]{\ensuremath{\mathbfl{ #1 }}}    
\newcommand{\rvecc}[2]{\ensuremath{
              \left( \begin{array}{c} \!\!#1\!\! \\
                                      \!\!#2\!\! \end{array} \right) }}
\newcommand{\tp}{\ensuremath{^\mathsf{T}}}

\begin{document}

   \title{Stereoscopy of extreme UV quiet Sun brightenings observed by Solar Orbiter/EUI}


   \author{A.~N.~Zhukov\inst{\ref{i:rob}, \ref{i:sinp}}\fnmsep\thanks{Corresponding author: Andrei Zhukov \email{Andrei.Zhukov@sidc.be}}
        \and
        M. Mierla \inst{\ref{i:rob}, \ref{i:igar}}
        \and
        F. Auch\`ere\inst{\ref{i:ias}}
        \and
        S. Gissot\inst{\ref{i:rob}}
        \and
        L. Rodriguez\inst{\ref{i:rob}}
        \and
        E. Soubri\'e\inst{\ref{i:ias},\ref{i:iaccc}}
        \and
        W.~T. Thompson\inst{\ref{i:gsfc}} 
        \and
        B.~Inhester \inst{\ref{i:mps}} 
        \and
        B.~Nicula\inst{\ref{i:rob}}
        \and
        P. Antolin\inst{\ref{i:north}}
        \and
        S. Parenti\inst{\ref{i:ias}}
        \and
        É. Buchlin\inst{\ref{i:ias}}
        \and
        K.~Barczynski\inst{\ref{i:pmod},\ref{i:eth}}
        \and
        C. Verbeeck\inst{\ref{i:rob}}
        \and
        E. Kraaikamp\inst{\ref{i:rob}}
        \and
        P.~J. Smith\inst{\ref{i:mssl}}
        \and
        K.~Stegen\inst{\ref{i:rob}}
        \and
        L. Dolla\inst{\ref{i:rob}}
        \and
        L.~Harra\inst{\ref{i:pmod},\ref{i:eth}}
        \and
        D.~M. Long\inst{\ref{i:mssl}}
        \and
        U. Sch\"uhle\inst{\ref{i:mps}}
        \and
        O. Podladchikova\inst{\ref{i:pmod}}
        \and
        R. Aznar Cuadrado\inst{\ref{i:mps}}
        \and
        L. Teriaca\inst{\ref{i:mps}}
        \and
        M.~Haberreiter\inst{\ref{i:pmod}}
        \and
        A.~C.~Katsiyannis\inst{\ref{i:rob}}
        \and
        P.~Rochus\inst{\ref{i:csl}}
       \and
        J.-P. Halain \inst{\ref{i:csl},\ref{i:esa}}
       \and
        L. Jacques\inst{\ref{i:csl}}
        \and
        D. Berghmans\inst{\ref{i:rob}} 
    }
    \institute{
            Solar-Terrestrial Centre of Excellence -- SIDC, Royal Observatory of Belgium, Ringlaan -3- Av. Circulaire, 1180 Brussels, Belgium\label{i:rob}
            \and
            Skobeltsyn Institute of Nuclear Physics, Moscow State University, 119992 Moscow, Russia\label{i:sinp}
            \and
            Institute of Geodynamics of the Romanian Academy, Bucharest, Romania\label{i:igar}
            \and
            Université Paris-Saclay, CNRS, Institut d'Astrophysique Spatiale, 91405, Orsay, France\label{i:ias}
            \and
            Institute of Applied Computing \& Community Code, Universitat de les Illes Balears, 07122 Palma de Mallorca, Spain\label{i:iaccc}
            \and
            Adnet Systems Inc., NASA Goddard Space Flight Center, Code 671, Greenbelt, MD 20771, United States of America\label{i:gsfc}
            \and
            Max Planck Institute for Solar System Research, Justus-von-Liebig-Weg 3, 37077 G\"ottingen, Germany\label{i:mps}
            \and
            Department of Mathematics, Physics and Electrical Engineering, Northumbria University, Newcastle Upon Tyne, NE1 8ST, United Kingdom\label{i:north}
            \and
            UCL-Mullard Space Science Laboratory, Holmbury St.\ Mary, Dorking, Surrey, RH5 6NT, UK\label{i:mssl}
            \and
            Physikalisch-Meteorologisches Observatorium Davos, World Radiation Center, 7260, Davos Dorf, Switzerland\label{i:pmod}
            \and
            ETH-Z\"urich, Wolfgang-Pauli-Str. 27, 8093 Z\"urich, Switzerland\label{i:eth}
            \and
            Centre Spatial de Li\`ege, Universit\'e de Li\`ege, Av. du Pr\'e-Aily B29, 4031 Angleur, Belgium\label{i:csl}
            \and
            European Space Agency (ESA/ESTEC),  Keplerlaan 1, PO Box 299 NL-2200 AG Noordwijk, The Netherlands\label{i:esa}
    }

   \date{Received April XXX, 2021; accepted April XXX, 2021}


  \abstract
    {The three-dimensional fine structure of the solar atmosphere is still not fully understood as most of the available observations are taken from a single vantage point.}
   {The goal of the paper is to study the three-dimensional (3D) distribution of the small-scale brightening events (``campfires'') discovered in the extreme-UV quiet Sun by the Extreme Ultraviolet Imager (EUI) aboard Solar Orbiter.}
   {We used a first commissioning data set acquired by the EUI's High Resolution EUV telescope (\hrieuv) on 30 May 2020 in the \SI{174}{\angstrom} passband and we combined it with simultaneous data taken by the Atmospheric Imaging Assembly (AIA) aboard the Solar Dynamics Observatory in a similar  \SI{171}{\angstrom} passband. The two-pixel spatial resolution of the two telescopes is 400~km and 880~km, respectively, which is sufficient to identify the campfires in both data sets. The two spacecraft had an angular separation of around 31.5$^\circ$ (essentially in heliographic longitude), which allowed for the three-dimensional reconstruction of the campfire position. These observations represent the first time that stereoscopy was achieved for brightenings at such a small scale. Manual and automatic triangulation methods were used to characterize the campfire data.}
   {The height of the campfires is located between 1000~km and 5000~km above the photosphere and we find a good agreement between the manual and automatic methods. The internal structure of campfires is mostly unresolved by AIA; however, for a particularly large campfire, we were able to triangulate a few pixels, which are all in a narrow range between 2500 and 4500~km.}
{We conclude that the low height of  EUI campfires suggests that they belong to the previously unresolved fine structure of the transition region and low corona of the quiet Sun. They are probably apexes of small-scale dynamic loops heated internally to coronal temperatures. This work demonstrates that high-resolution stereoscopy of structures in the solar atmosphere has become feasible.}


\maketitle

%

\section{Introduction}

The solar atmosphere is remarkably inhomogeneous. Its fine structure reflects spatial scales on which various physical processes of energy storage and release take place. The crucial processes dominating the physics of the transition region and corona must be operating at very small spatial and temporal scales \citep[e.g.,][]{Parker1988, Klimchuk2015}. Processes taking place on different scales may be operational at the same time and the fine structure is often noticeably anisotropic, with observed phenomena such as loops, spicules, and plumes having a clear three-dimensional (3D) aspect. This suggests that knowledge of 3D geometry is important in improving the understanding of the physical mechanisms at work in this region. 

 Remarkable success has been achieved in recent years with regard to the 3D reconstruction of the global corona \citep[][]{Frazin2010, Vasquez2011} and different solar structures \citep[see, e.g., the review by ][]{Aschwanden2011LRSP}, such as loops \citep[e.g.,][]{Feng2007, Rodriguez2009, Aschwanden2011JASTP}, polar plumes \citep{Barbey2008, Feng2009, Barbey2011}, prominences \citep[e.g.,][]{Gissot2008, Bemporad2009, Zhou2021}, loops with coronal rain \citep{Pelouze_2021}, streamers \citep[e.g.,][]{Saez2005, Wang2007, Zhukov2008, Sasso2019}, and coronal mass ejections \citep[e.g.,][]{Mierla2010, Frazin2012}. The two-spacecraft STEREO mission \citep[Solar-TErrestrial RElations Observatory, see ][]{STEREO} was the first mission dedicated to the stereoscopic observations of the solar atmosphere. However, the 3D structure of the solar atmosphere at high resolution has not been sufficiently explored. The 3D reconstruction of the solar atmosphere by STEREO is limited by the spatial resolution of the EUVI telescopes \citep[Extreme-UltraViolet Imager, see ][]{SECCHI}, which is \SI{3.2}{\arcsecond} for two pixels, or \SI{2300}{\kilo\meter} at \SI{1}{\astronomicalunit}. The highest resolution of extreme-ultraviolet (EUV) observations of the corona and upper transition region was reached by the Hi-C sounding rocket \citep{Cirtain2013, Rachmeler2019}: \SI{0.3}{\arcsecond}, or \SI{218}{\kilo\metre}. However, the Hi-C data could not be combined with the STEREO data as both STEREO spacecraft were facing the far side of Sun, and the distance between Hi-C and near-Earth telescopes such as the Atmospheric Imaging Assembly \citep[AIA, see][]{AIA} aboard the Solar Dynamics Observatory \citep[SDO, see ][]{SDO} is too small to provide sufficiently varied vantage points.
 
The Solar Orbiter mission \citep{SolarOrbiter} provides an additional and rapidly changing vantage point for observations of the solar atmosphere that can be combined with EUVI aboard the remaining STEREO~A spacecraft or with near-Earth telescopes such as SDO/AIA. An important advantage of Solar Orbiter is its close approach to the Sun, which results in the increase of the spatial resolution of its remote-sensing instruments. The Extreme Ultraviolet Imager \citep[EUI, ][]{EUI} aboard Solar Orbiter provides an improved spatial resolution of EUV observations in comparison with routinely available images from other telescopes. In particular, EUI's High Resolution Imager (HRI) operating at EUV wavelengths (\hrieuv) around 174~\AA\ has a two-pixel spatial resolution of \SI{0.984}{\arcsecond}, which corresponds to around 400~km near the first Solar Orbiter perihelion at the distance of \SI{0.556}{\astronomicalunit}. The \hrieuv passband is dominated by the emission of \mbox{Fe IX} and \mbox{Fe X} lines formed in the corona and in the upper transition region at temperatures around 0.6--1~MK. 

\begin{figure}[ht]
\centering
\includegraphics[width=1.0\hsize]{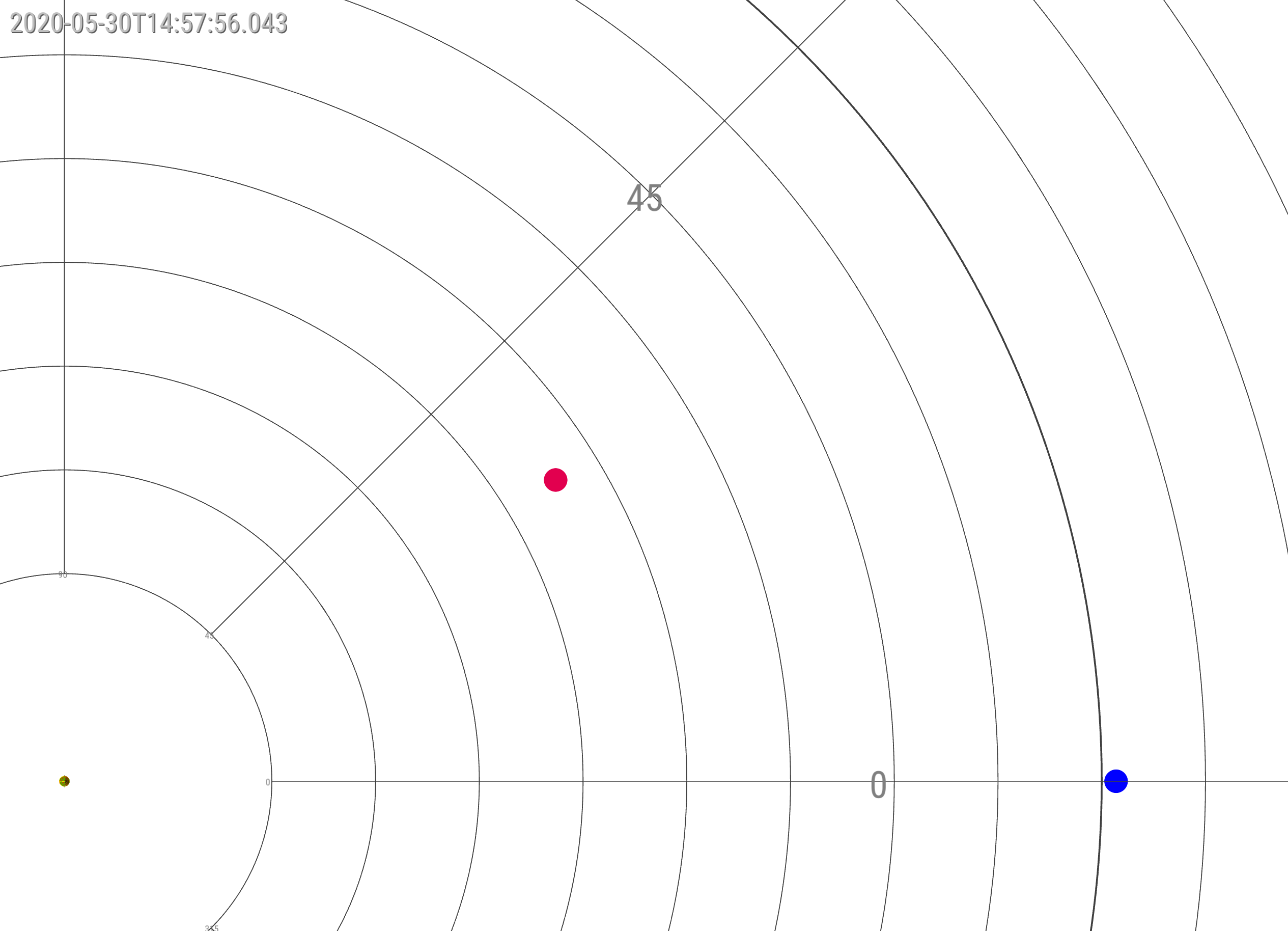}
\caption{Positions of the Solar Orbiter and SDO spacecraft (red and blue dots respectively) on 30 May 2020. The separation between the two spacecraft in the heliographic longitude is 31.495$^\circ$, and the latitudinal separation is $1.159^\circ$ (not shown), giving the true angular separation of 31.514$^\circ$. The thickest arc denotes the Earth's orbit, and the yellow dot is the Sun. The image is created using the JHelioviewer software \citep{JHelioviewer}.}
\label{fig:positions}
\end{figure}

The first operations of \hrieuv revealed a new type of small-scale EUV brightenings in the quiet Sun, dubbed ``campfires'' \citep{Berghmans2021}. These are structures between 400~km and 4000~km in size and between 10~s and 200~s in duration, located predominantly at the chromospheric network, as suggested by simultaneous images taken by EUI's second high-resolution telescope in the H~I Lyman-$\alpha$ passband (\hrilya). The differential emission measure (DEM) of campfires peaks at coronal temperatures around $\log T = 6.1$.
In the present work, we report the first stereoscopy of EUI campfires based on Solar Orbiter and SDO observations, describing in detail the campfire triangulation reported briefly by \citet{Berghmans2021}. The paper is organized as follows. In Sect.~\ref{section:Observations}, we present the observational data. Sections~\ref{section:manual} and \ref{section:automatic} present the triangulations made independently by manual and automatic methods, respectively, and a comparison of the results. Our discussion and conclusions are given in Sect.~\ref{section:Discussion}.

\section{Observations}
\label{section:Observations}

\begin{figure*}[ht]
\centering
{ \includegraphics[width=1.0\hsize]{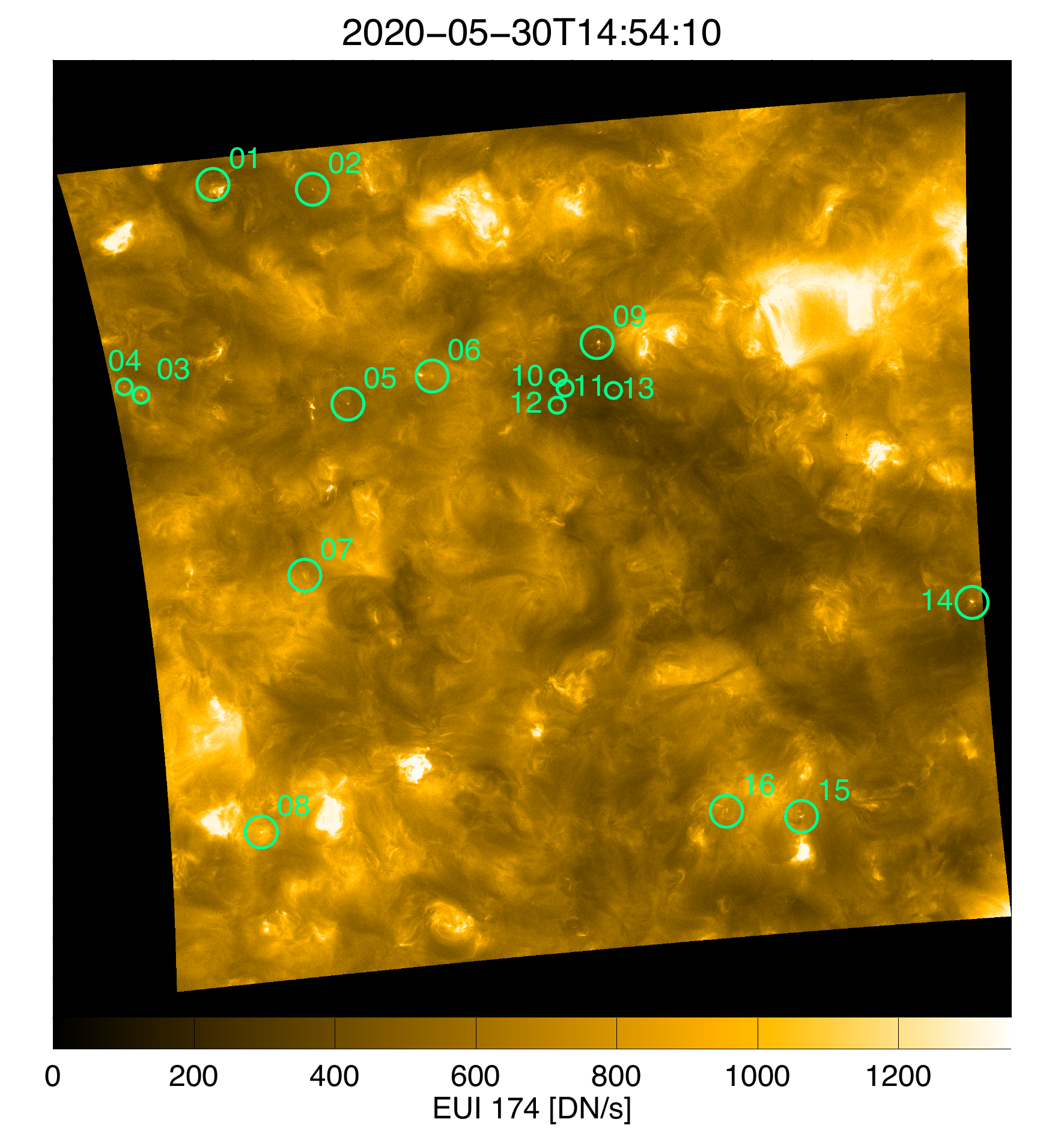}}
\caption{Solar Orbiter/EUI \hrieuv image taken on 30 May 2020 at 14:54:10~UT. The image was remapped to Carrington coordinates to facilitate comparison with Figs.~\ref{fig:aia} and \ref{fig:hmi}. The sixteen campfires selected for the manual triangulation are shown with green circles and labeled with their respective numbers. Solar north is on top, west is to the right. 
}
\label{fig:fov}
\end{figure*}

\begin{figure*}[ht]
\centering
{ \includegraphics[width=1.0\hsize]{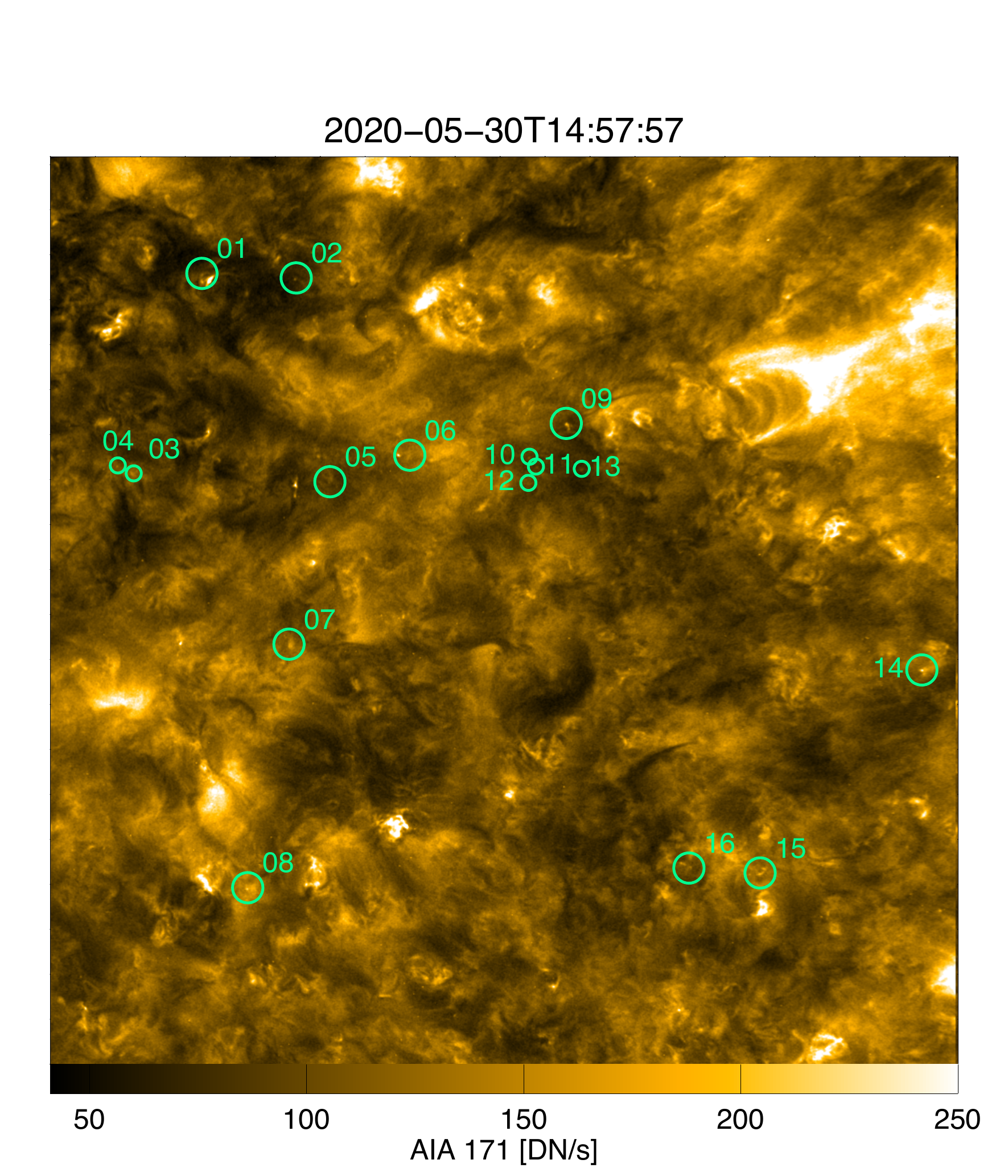}}
\caption{SDO/AIA 171~\AA\ image taken on 30 May 2020 at 14:57:57~UT.
The field of view is the same as that of Fig.~\ref{fig:fov}.
}
\label{fig:aia}
\end{figure*}

This work uses release 1 of properly calibrated and formatted EUI data\footnote{DOI: \url{https://doi.org/10.24414/wvj6-nm32}. We note that the EUI data from release 2 for the same period do not differ in a way significant for this study.}, as described in \citet{Berghmans2021}. Fifty images were taken on 30 May 2020 between 14:54:00~UT and 14:58:05~UT at the cadence of 5~s and two-pixel spatial resolution of around 400~km. At that moment, Solar Orbiter was situated at 0.556~au from the Sun and \ang{31.5} to the west from the Earth (Fig.~\ref{fig:positions}). See \citet{Berghmans2021} for a detailed description of the acquired \hrieuv data set. 

\begin{figure*}[ht]
\centering
{ \includegraphics[width=1.0\hsize]{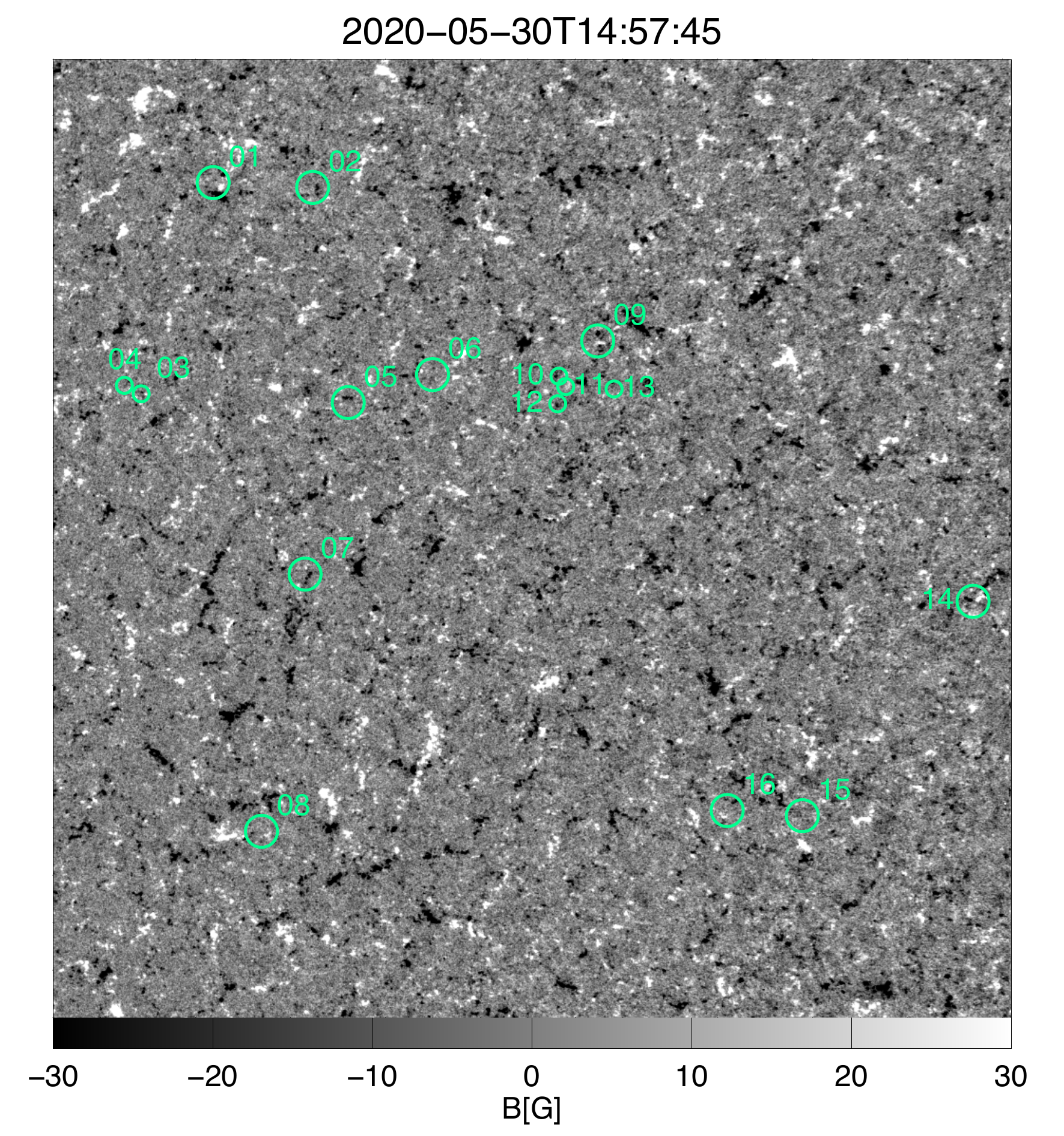}}
\caption{Line-of-sight magnetogram taken by SDO/HMI on 30 May 2020 at 14:57:37~UT. This image and the images shown in Figs.~\ref{fig:fov} and \ref{fig:aia} are nearly simultaneous representations of the same solar scene, due to different light travel times from the Sun to the two observing spacecraft. The image was remapped to Carrington coordinates and the field of view is the same as that of Fig.~\ref{fig:fov}. The sixteen campfires are shown with green circles and labeled with their numbers. }
\label{fig:hmi}
\end{figure*}

To perform the stereoscopy, we supplemented the \hrieuv images with the data acquired in a similar 171~\AA\ passband by the AIA telescope \citep[Atmospheric Imaging Assembly, ][]{AIA}, which takes solar images at \SI{1.5}{\arcsecond} resolution from the Solar Dynamics Observatory mission \citep[SDO, ][]{SDO} situated at that time at \SI{1.014}{\astronomicalunit} from the Sun (in Earth's orbit). For the manual stereoscopy (Sect.~\ref{section:manual}), we chose the image taken by \hrieuv at 174~\AA\  at 14:54:10~UT, which displays a number of small and bright campfires (Fig.~\ref{fig:fov}). The corresponding AIA 171~\AA\ image was acquired at 14:57:57~UT (see Fig.~\ref{fig:aia}). Given the different light travel time from the Sun to each spacecraft, the two telescopes image the same scene simultaneously to within 2~s. For the automatic stereoscopy (Sect.~\ref{section:automatic}), we used 50 pairs of nearly simultaneous \hrieuv and AIA images,  selected, once again, by taking into account the light travel time to each spacecraft. 

To investigate the relation between the campfires and underlying photospheric magnetic field, we used the photospheric line-of-sight magnetogram obtained with the Helioseismic and Magnetic Imager \citep[HMI,][]{HMI} aboard SDO. The HMI provides the magnetic field data at \SI{1.0}{\arcsecond} spatial resolution. For our analysis, we choose the magnetogram taken at 14:57:45~UT (Fig.~\ref{fig:hmi}), which is nearly simultaneous to the HRI and AIA images shown in Figs.~\ref{fig:fov} and \ref{fig:aia}. To facilitate the comparison, the three images are remapped to Carrington coordinates \citep{Berghmans2021}. The preliminary analysis suggests the campfires appear at the magnetic network. Seven campfires (03, 06, 07, 09, 12, 13, 14) are clearly projected on tiny photospheric bipoles. Three campfires (04, 10, 11) are projected on single polarity patches which do not belong to clear bipoles. Six campfires (01, 02, 05, 08, 15, 16) are projected on areas with very weak magnetic field (below 10~G). 

\begin{figure*}[ht]
\centering
{ \includegraphics[width=1.0\hsize]{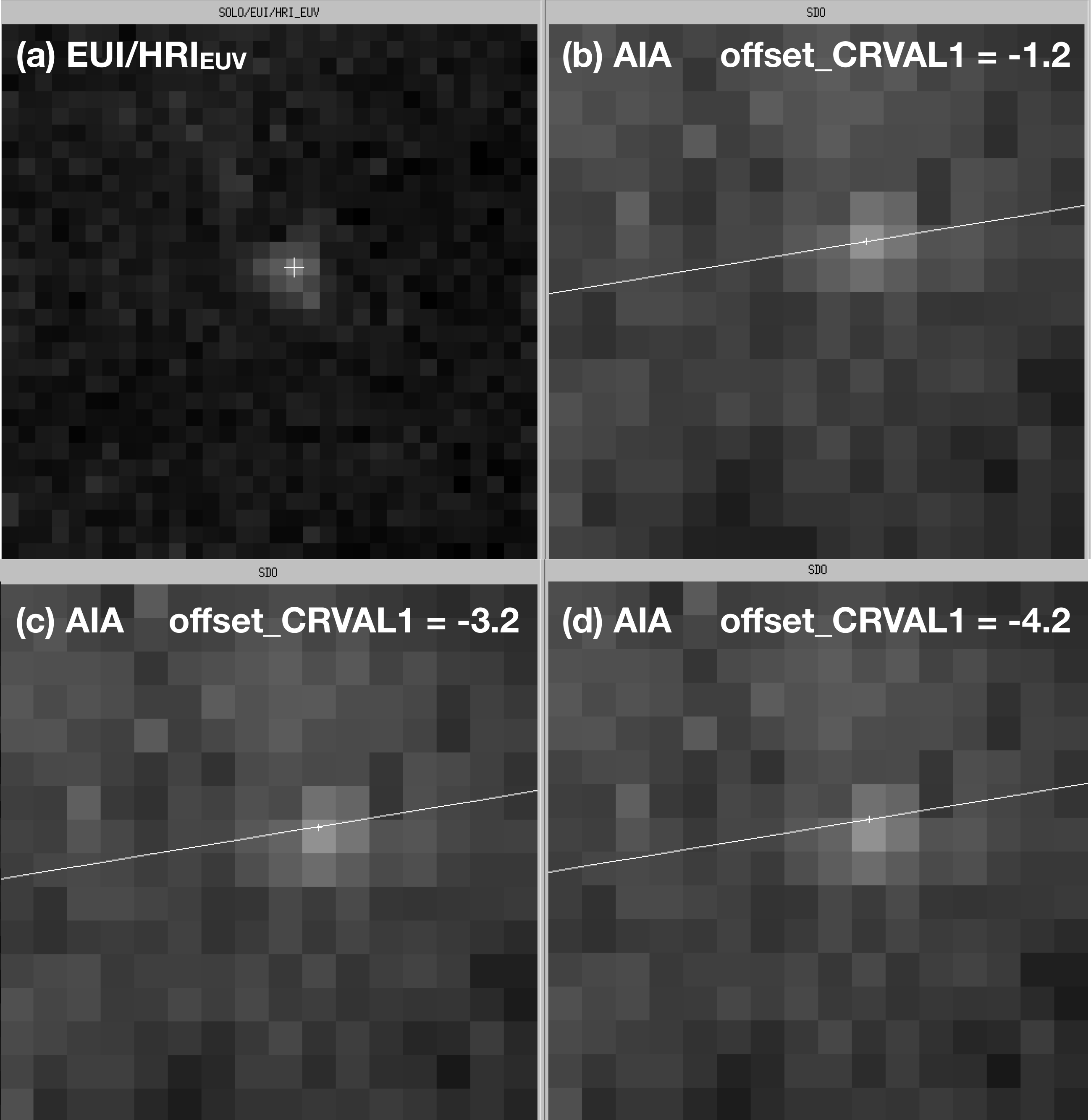}} 
\caption{Example of using the ssc\_measure software for the manual triangulation of campfire 12 (see Fig.~\ref{fig:fov}). Panel (a) shows a part of the \hrieuv image taken on 30 May 2020 at 14:54:10~UT. The cross marks the central and the brightest pixel of the campfire. The other three panels show a part of the nearly simultaneous AIA 171~\AA\ image with the epipolar line corresponding to the cross in panel (a) overplotted in white for the offsets of the CRVAL1 value of \SI{-1.2}{\arcsecond} (b), \SI{-3.2}{\arcsecond} (c), and \SI{-4.2}{\arcsecond} (d). The offset of the CRVAL2 value is \SI{-16.7}{\arcsecond}. The counterpart of the \hrieuv pixel marked with the cross must be on the epipolar line in the AIA image.  
}
\label{fig:manual}
\end{figure*}

\section{Manual triangulation}
\label{section:manual}

\begin{figure*}[ht]
\centering
\begin{subfigure}[t]{0.45\textwidth}
\centering
{ \includegraphics[width=1.0\hsize]{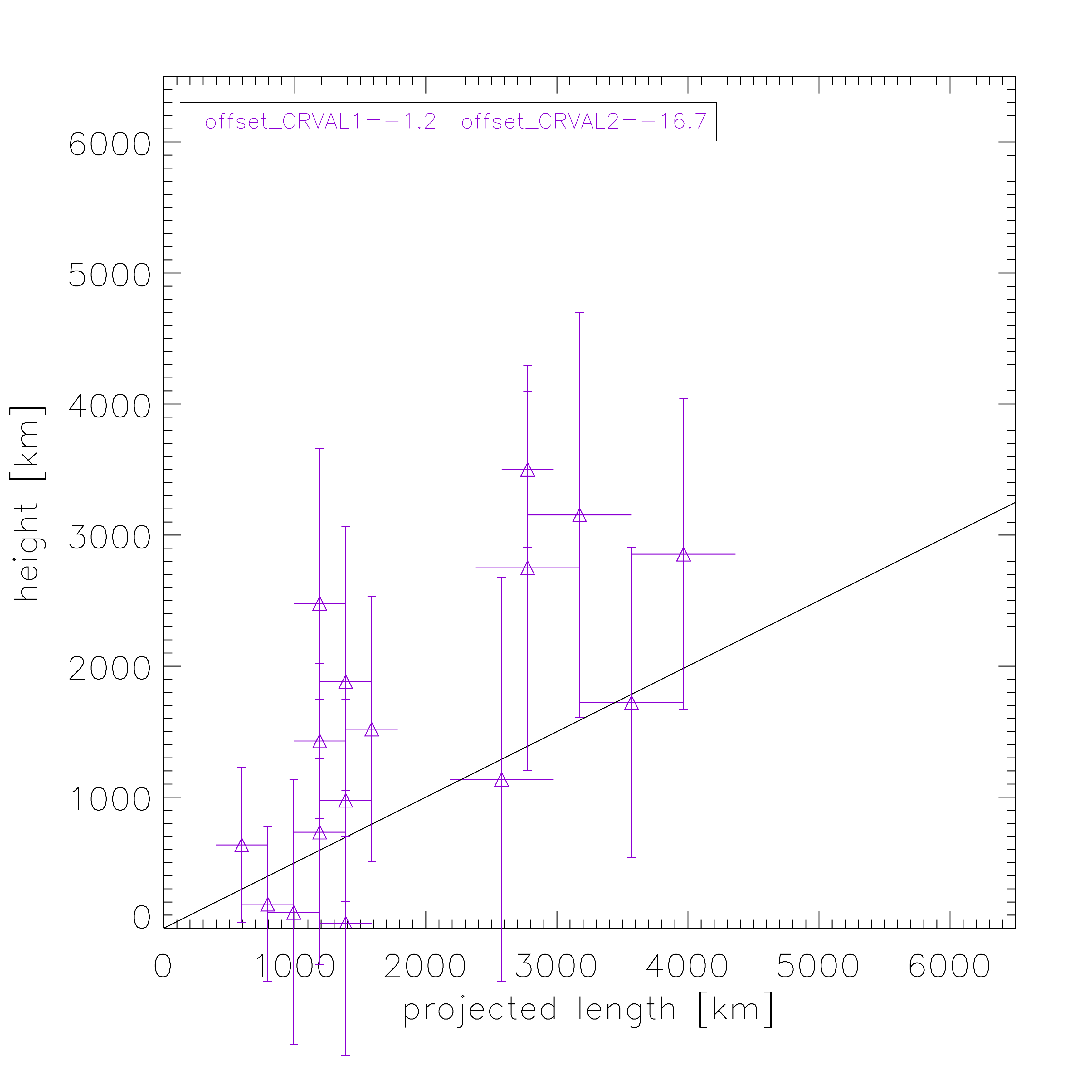}}
\end{subfigure}
\begin{subfigure}[t]{0.45\textwidth}
\centering
{ \includegraphics[width=1.0\hsize]{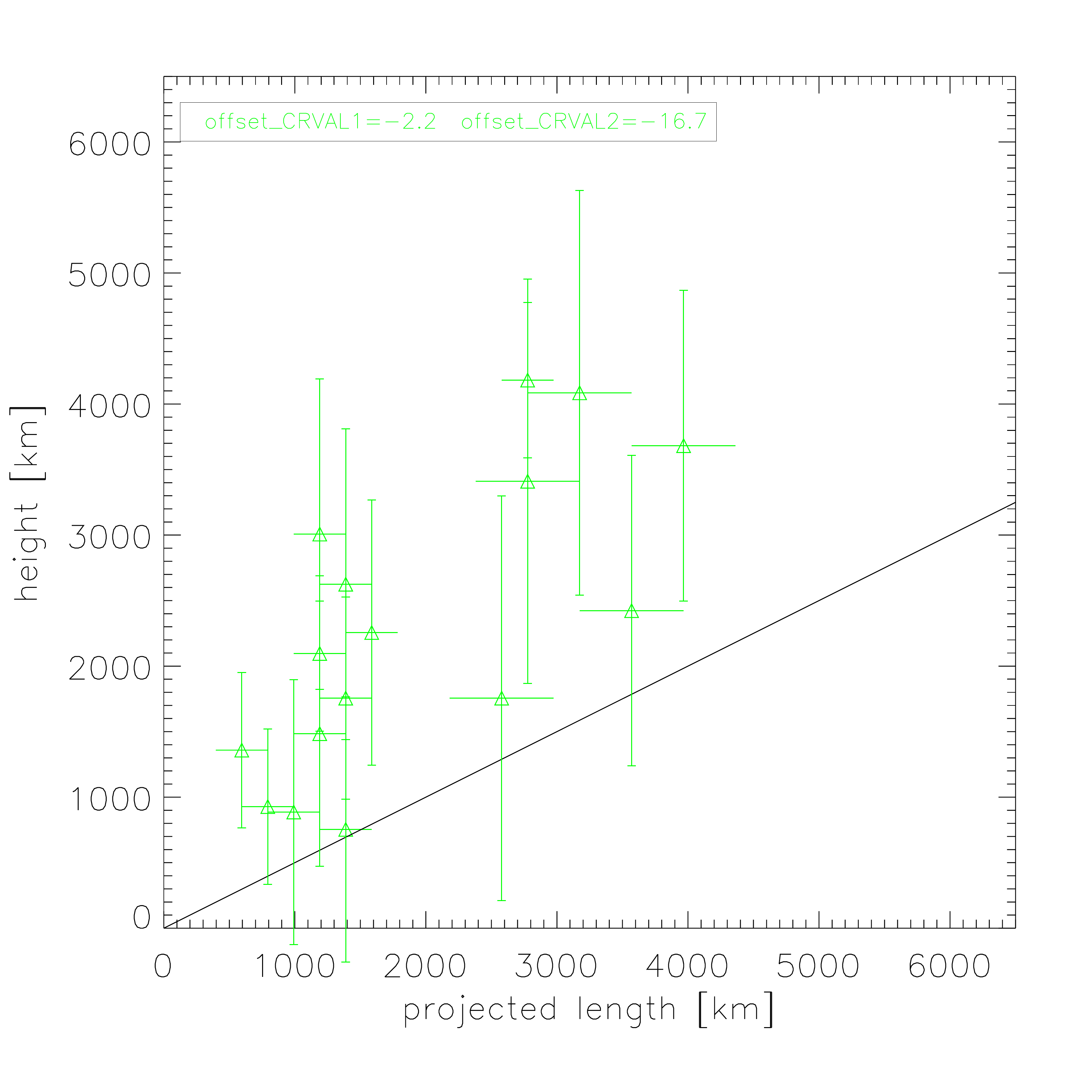}}
\end{subfigure}

\begin{subfigure}[t]{0.45\textwidth}
\centering
{ \includegraphics[width=1.0\hsize]{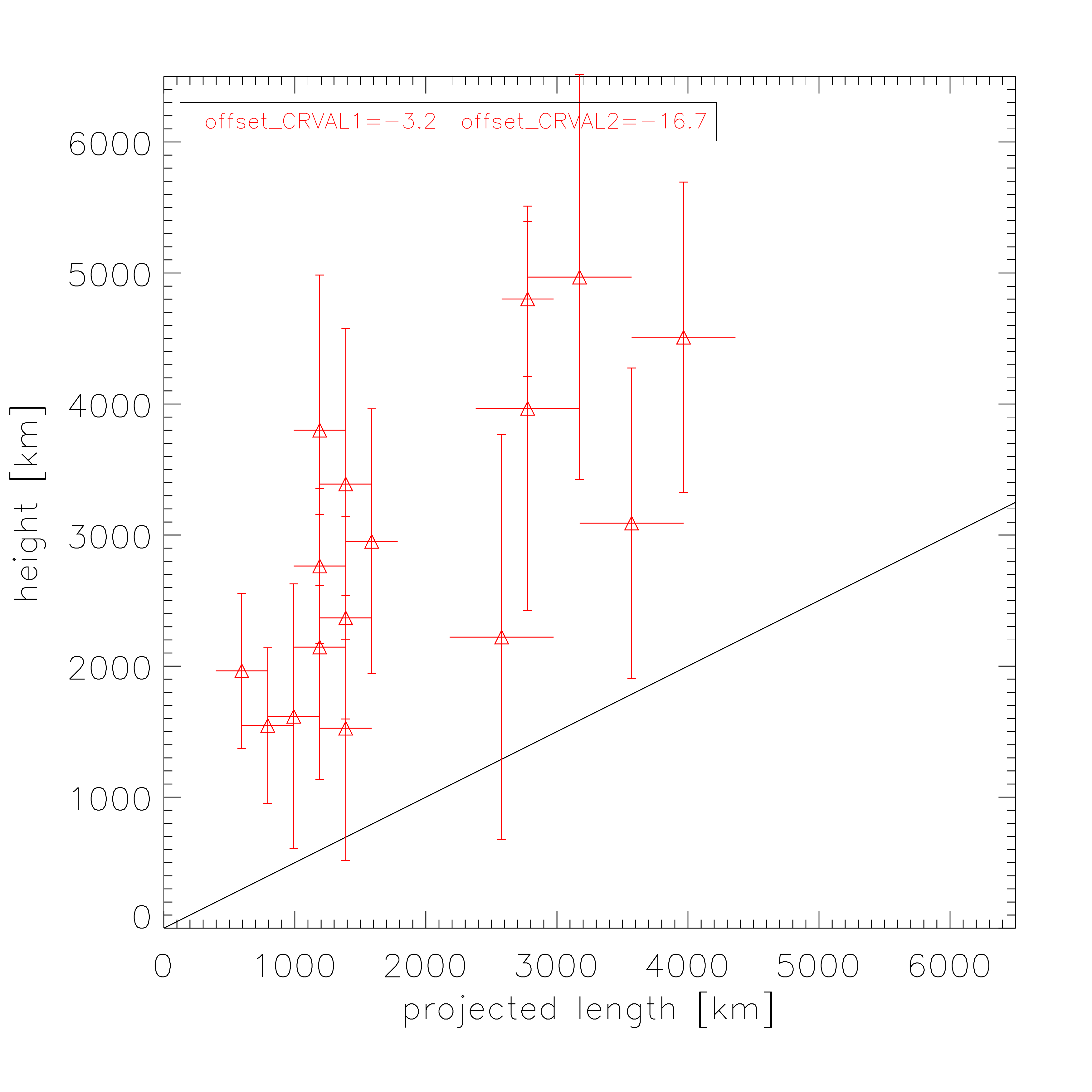}}
\end{subfigure}
\begin{subfigure}[t]{0.45\textwidth}
\centering
{ \includegraphics[width=1.0\hsize]{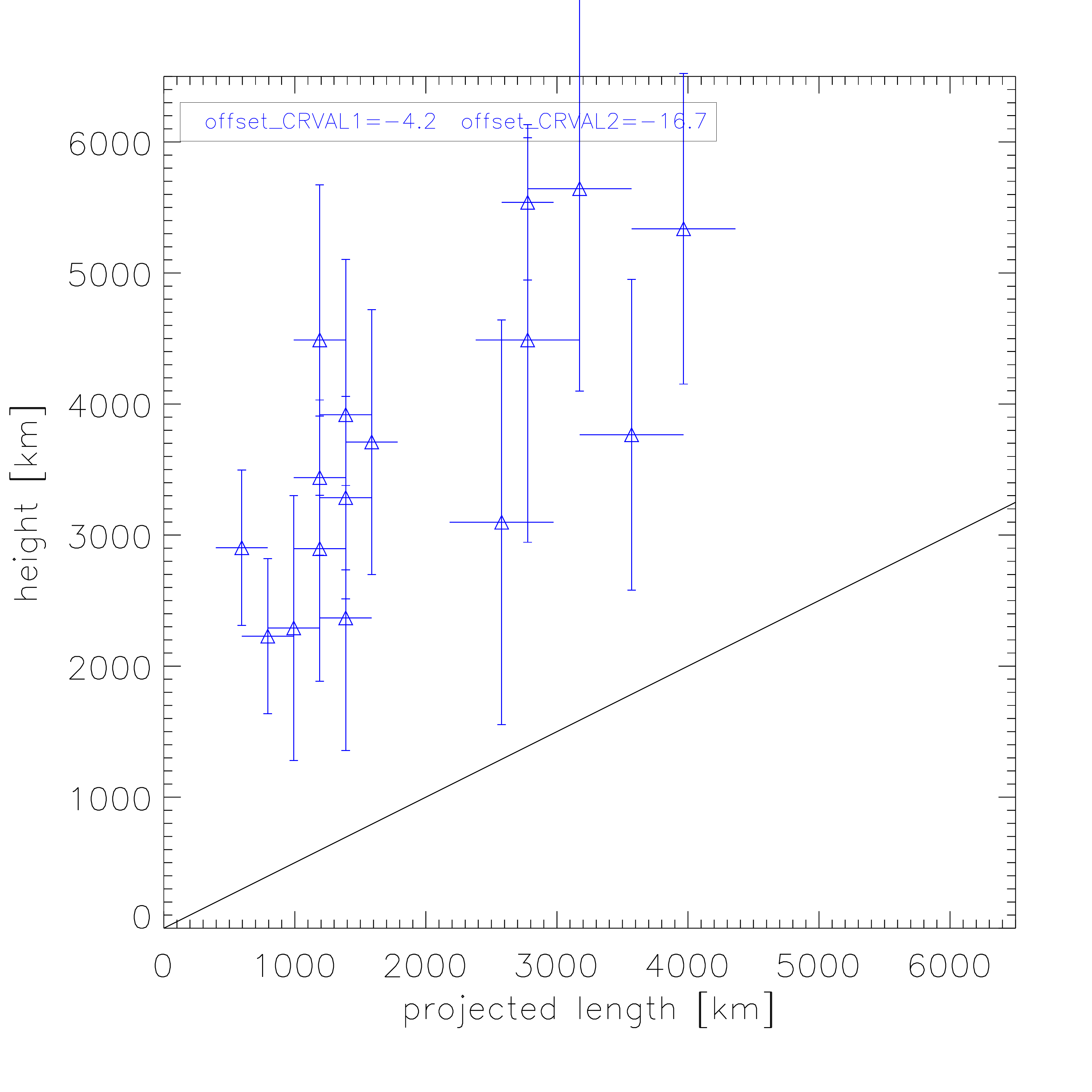}}
\end{subfigure}
\caption{
Height above the photosphere of the 16 campfires measured using the manual method versus their projected length (triangles with their error bars, $h \pm \delta h$). The \hrieuv image taken at 14:54:10~UT is used. 
Each panel shows the triangulation results for different values of the CRVAL1 offset.
The black line in each panel represents the height corresponding to the half of the projected length, as we would expect for a semi-circular loop, for comparison.}
\label{fig:height_size}
\end{figure*}

We first determined the height of campfires above the photosphere using a manual triangulation. Inspecting the images by eye, we selected sixteen small-size campfires that could be clearly seen in both \hrieuv (14:54:10~UT) and AIA (14:57:57~UT) images, see Fig.~\ref{fig:fov} and Table~\ref{table:1}. 

\subsection{Method of manual triangulation}
\label{section:method_manual}

\begin{table*}
\caption{\label{table:1}Manual triangulation of sixteen campfires.}
\centering 
\begin{tabular}{c c c c c c c c}
\hline\hline
Number & $\varphi$, degrees & $\theta$, degrees & $h$, km & $x_{\rm AIA}$ & $y_{\rm AIA}$ & $x_{\rm HRI}$ & $y_{\rm HRI}$ \\
\hline 
01 &   11.25320 &   22.45690 &    3100 $\pm$ 1000 &  2333.94043 & 2675.23169 &  325.87500 & 1986.00000 \\
02 &  15.27350  &  22.27340  &   1700 $\pm$ 600 & 2434.00293 & 2668.92017 &  543.12500 & 1949.50000 \\
03 &    8.27509 &   13.87120 &    1700 $\pm$ 1000 & 2268.93018 & 2449.54443 &   62.87500 & 1494.87500 \\
04 &    7.60888 &   14.21930 &    2200 $\pm$ 600 & 2250.99951 & 2459.15894 &   29.43750 & 1520.50000 \\
05 &   16.71790 &   13.52400 &    2300 $\pm$ 1000 & 2491.44409 & 2439.83472 &  536.00000 & 1422.37500 \\
06 &   20.14650 &   14.66880 &    1800 $\pm$ 1000 & 2576.00269 & 2469.73364 &  745.93750 & 1470.87500 \\
07 &   14.99780 &    6.48183 &    4700 $\pm$ 1200 & 2457.04565 & 2249.04028 &  375.75000 & 1003.50000 \\
08 &   13.19160 &   -3.93560 &    2570 $\pm$ 800 & 2408.98730 & 1960.20178 &  214.50000 &  373.25000 \\
09 &   26.92540 &   16.02440 &    2400 $\pm$ 1500 & 2738.42334 & 2505.08618 & 1154.62500 & 1515.62500 \\
10 &   25.30190 &   14.57280 &    3300 $\pm$ 1200 & 2704.99585 & 2467.24341 & 1048.50000 & 1437.50000 \\
11 &    25.59230 &    14.14220 &    4000 $\pm$ 1200 & 2713.99805 & 2456.07397 & 1062.62500 & 1410.43750 \\
12 &   25.26470 &   13.46310 &     3000 $\pm$ 600 & 2706.97290 & 2437.28711 & 1039.00000 & 1370.00000 \\
13 &   27.55310 &   14.05250 &    3600 $\pm$ 1200 & 2761.06250 & 2453.06201 & 1179.37500 & 1394.00000 \\
14 &  42.09840  &   5.42490 &    4200 $\pm$ 1500 & 3109.29858 & 2214.52197 & 2020.87500 &  792.75000 \\
15 &   35.12160 &   -3.30069 &    5200 $\pm$ 1500 & 2962.86011 & 1974.08875 & 1550.00000 &  285.25000 \\
16 &   32.13900 &   -3.03113 &    5000 $\pm$ 600 & 2893.99707 & 1982.23596 & 1367.00000 &  316.93750 \\
\hline
\end{tabular}
\tablefoot{
First column: campfire number. Second and third columns: Stonyhurst longitude and latitude. Fourth column: height above the photosphere (for the CRVAL1 and CRVAL2 offsets of \SI{-3.2}{\arcsecond} and \SI{-16.7}{\arcsecond} respectively). Fifth and sixth columns: pixel coordinates of the triangulated point in the AIA image. Seventh and eighth columns: pixel coordinates of the triangulated point in the HRI image.
}
\end{table*}

The triangulation method requires identification of the same point in the two images \citep[a process called tie-pointing; see, e.g., ][]{Inhester2006}. The 3D positions of lines of sight (LOS) passing through the point that is visible in the two images are calculated, and the position of the intersection point in three-dimensional space is determined; for more details, see \citet{Inhester2006} and Appendix~\ref{app}. We performed the triangulation using the scc\_measure.pro program of SolarSoft \citep{Thompson2008, ThompsonW2009}, which outputs the position of campfires in Stonyhurst coordinates (longitude, latitude, and height from the Sun's center, see \citeauthor{Thompson2006} \citeyear{Thompson2006}). Then we subtract the photospheric radius of 695700~km \citep{Prsa2016} to get the height, $h$,  above the photosphere. Figure~\ref{fig:manual} illustrates the process of using the ssc\_measure.pro program. We select the center of the brightest part of the campfire in the \hrieuv image. In the case of campfire 12, shown in Fig.~\ref{fig:manual}, this is the center of the brightest pixel closest to the center of the campfire. In case there are a few adjacent pixels of essentially the same brightness, we select the geometric center of the pixels. The software calculates the corresponding epipolar line \citep{Inhester2006} and projects it on the AIA image. The brightest AIA campfire pixel on the epipolar line is selected as the AIA pixel corresponding to the \hrieuv pixel. The center of the epipolar line interval crossing the selected AIA pixel is used for triangulation. The 3D position of the crossing of the lines of sight passing through the selected points in the \hrieuv and AIA images is calculated. 

\subsection{Errors of manual triangulation}
\label{section:errors}

The extensions along the corresponding lines of sight of selected \hrieuv and AIA pixels intersect to define the volume in 3D space where the emitting material is situated. The vertical extent of this volume defines the error of the triangulated height. It can be determined from simple geometric considerations by extending the analysis reported by \citet{Inhester2006} to the condition of different pixel sizes of the two images\footnote{Equation~(3) in \citet{Rodriguez2009} that uses the tangent of the half-angle between the two vantage points instead of its sine (used in Fig.~11 of \citet{Inhester2006}) is valid only in the small-angle approximation.}. The half-error $\delta h$ depends on the linear pixel sizes of \hrieuv ($\delta s_1 = 198$~km) and AIA ($\delta s_2 = 441$~km) at the time of observation:
\begin{equation}
\delta h = \frac{\sqrt{\delta s_1^2 + \delta s_2^2 + 2 \, \delta s_1 \, \delta s_2 \, \cos \gamma}} {2 \sin \gamma},
\label{equation:error}
\end{equation}
where $\gamma = 31.514^\circ$ is the angular separation between Solar Orbiter and SDO. If the triangulated point (the center of a campfire) can be localized to within one pixel in both \hrieuv and AIA images, then the error is $\delta h \approx \pm 600$~km. This is the minimal random error originating from the limited spatial resolution of the telescopes. Localizing the triangulated point to one pixel is not always possible as campfires may have extended bright cores. For some of the campfires, the error in the image plane could be up to four \hrieuv pixels and two AIA pixels. According to Eq.~(\ref{equation:error}), in this case the random error would be around $\pm 1500$~km. 

\begin{figure*}[ht]
\centering
\begin{subfigure}[t]{0.7\textwidth}
\centering
{ \includegraphics[width=1.0\hsize]{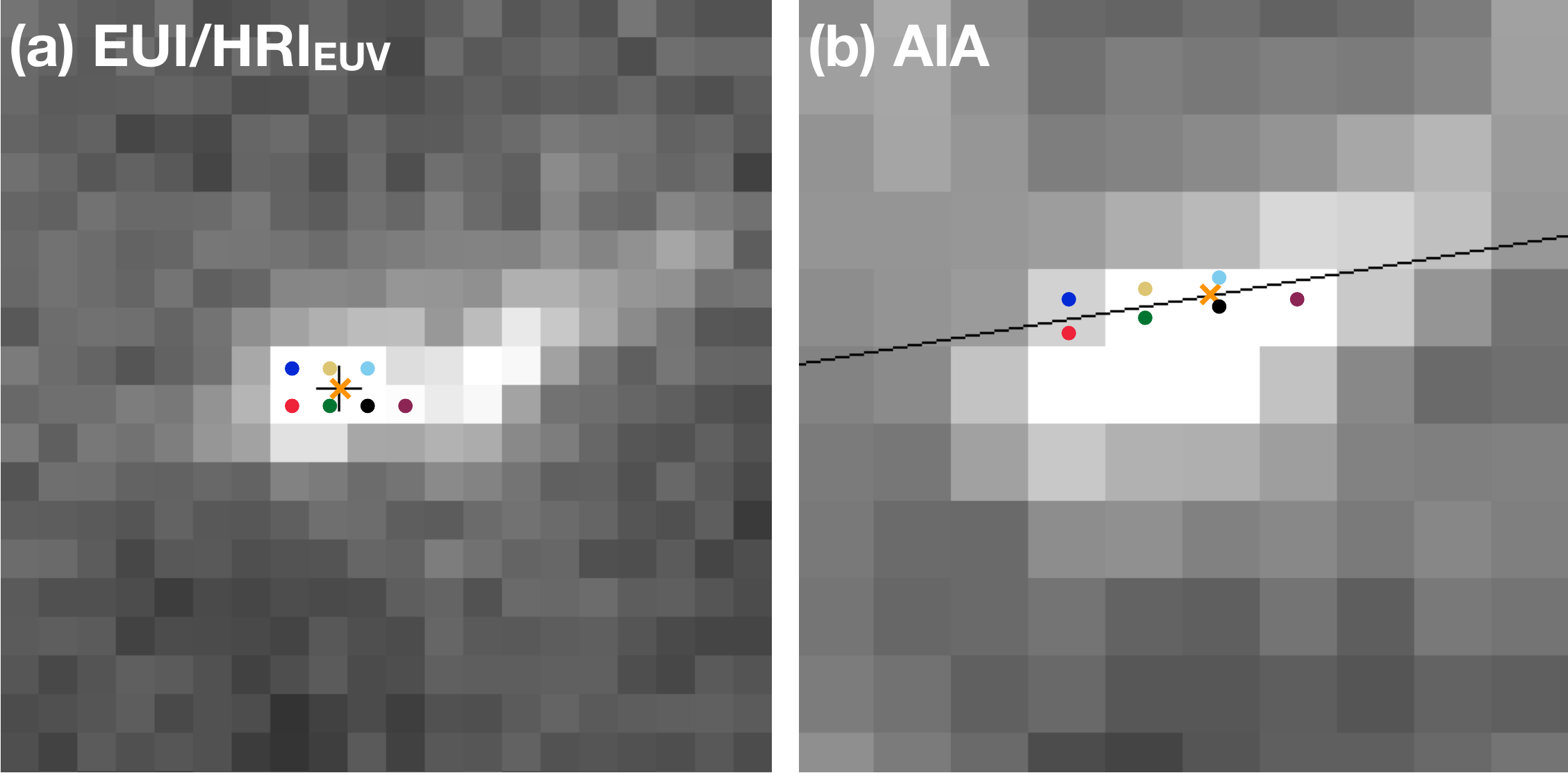}}
\end{subfigure}

\begin{subfigure}[t]{0.7\textwidth}
\centering
{ \includegraphics[width=1.0\hsize]{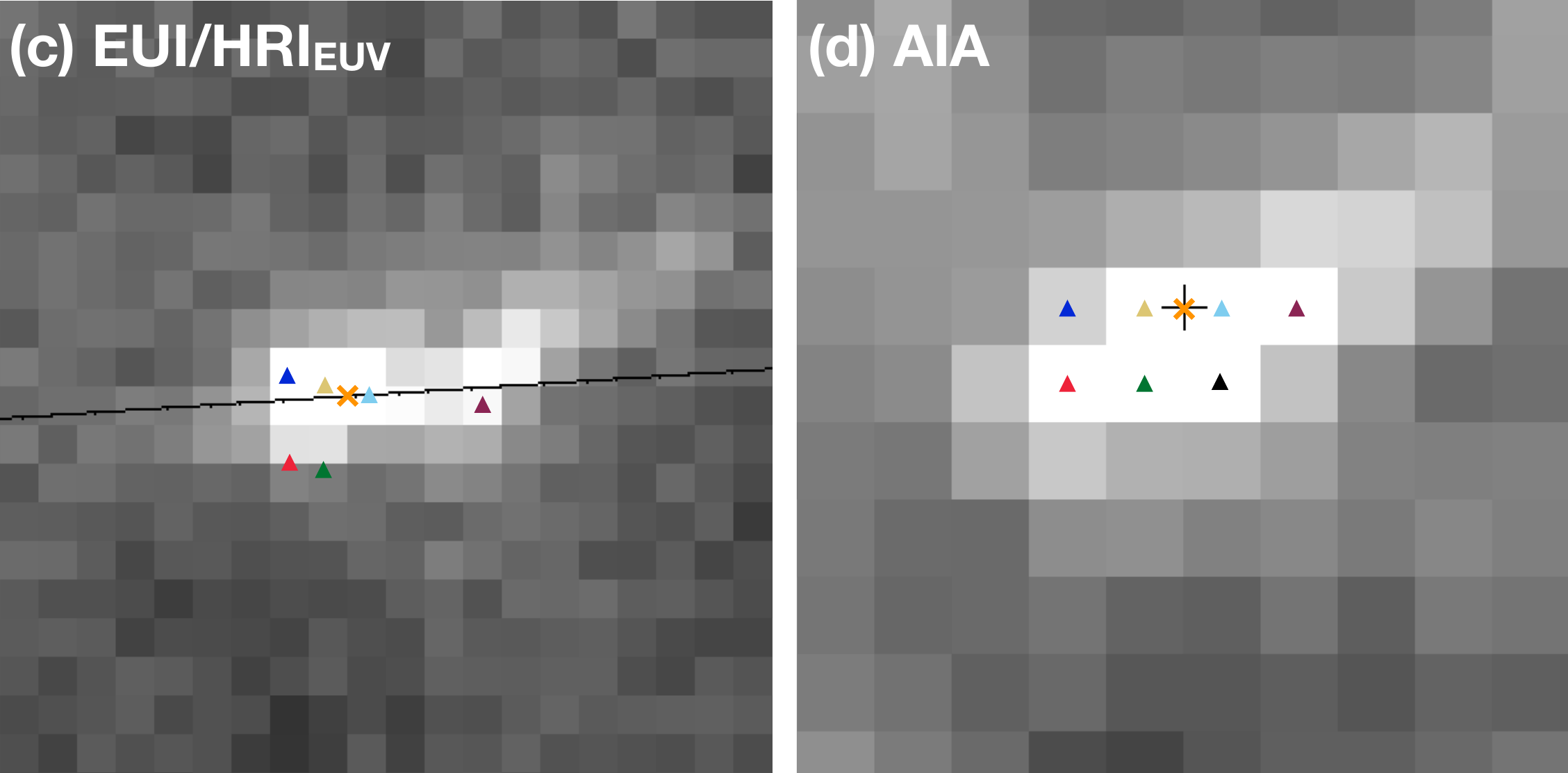}}
\end{subfigure}

\begin{subfigure}[t]{0.7\textwidth}
\centering
{ \includegraphics[width=1.0\hsize]{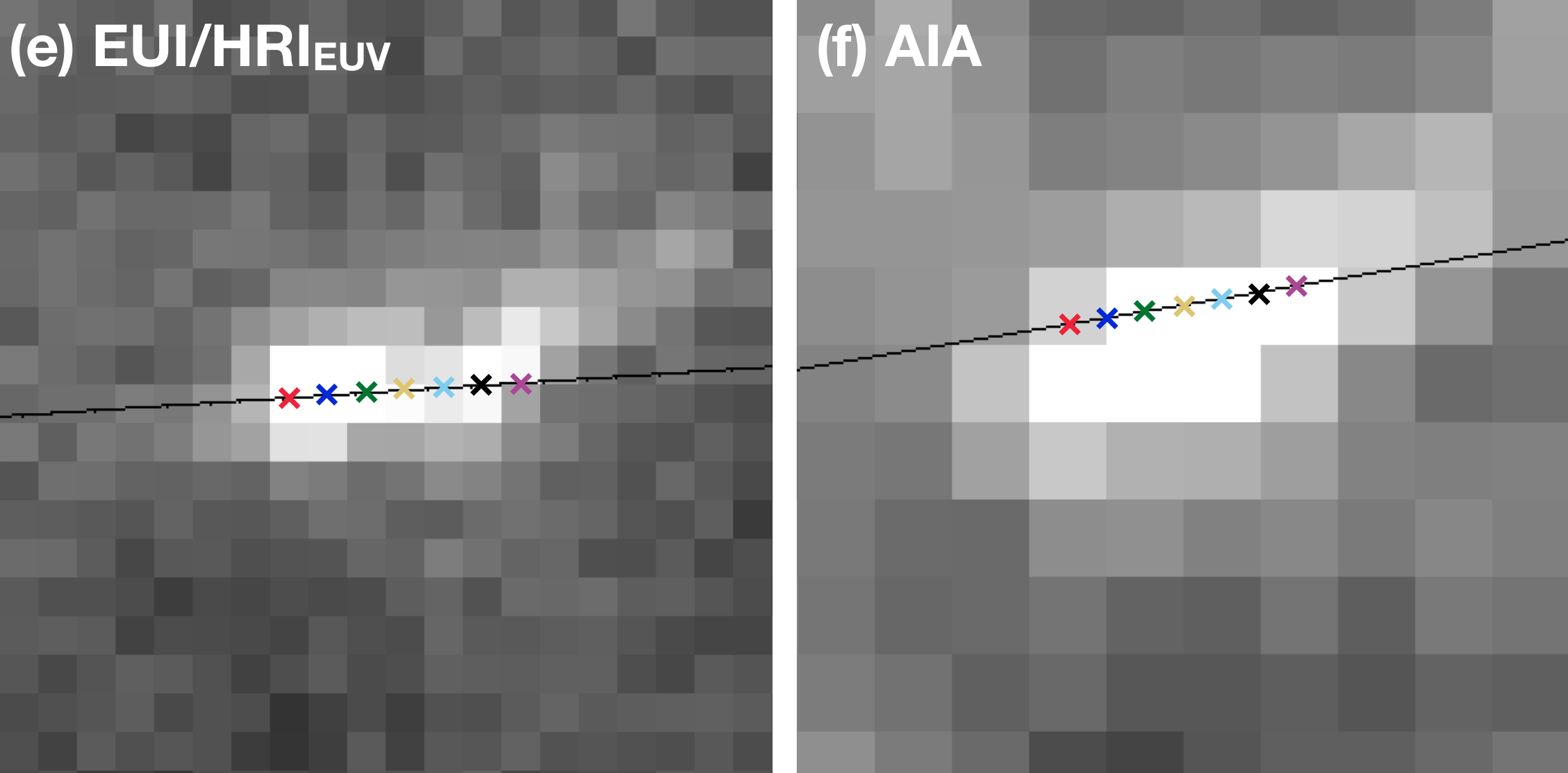}}
\end{subfigure}
\caption{
Triangulation of the fine structure of the campfire 15. Left panels: \hrieuv image. Right panels: SDO/AIA 171~\AA\ image. Top panels: Triangulation starting from the Solar Orbiter/\hrieuv data. The centers of each campfire pixel used for the triangulation are marked with colored dots in the top left panel. The dots of the same color mark the corresponding points in the AIA image (top right panel). The orange cross corresponds to the point for which the height is plotted in Fig.~\ref{fig:height_size}. The corresponding epipolar line is shown in white. Middle panels: same as top panels but starting from the SDO/AIA 171~\AA\ data. The bright pixel marked with the black triangle has no bright counterpart in the \hrieuv data. Bottom panels: Triangulation along an epipolar line (black lines).}
\label{fig:triang_hri}
\end{figure*}

\begin{figure*}[ht]
\centering
\begin{subfigure}[t]{0.33\textwidth}
\centering
{ \includegraphics[width=1.0\hsize]{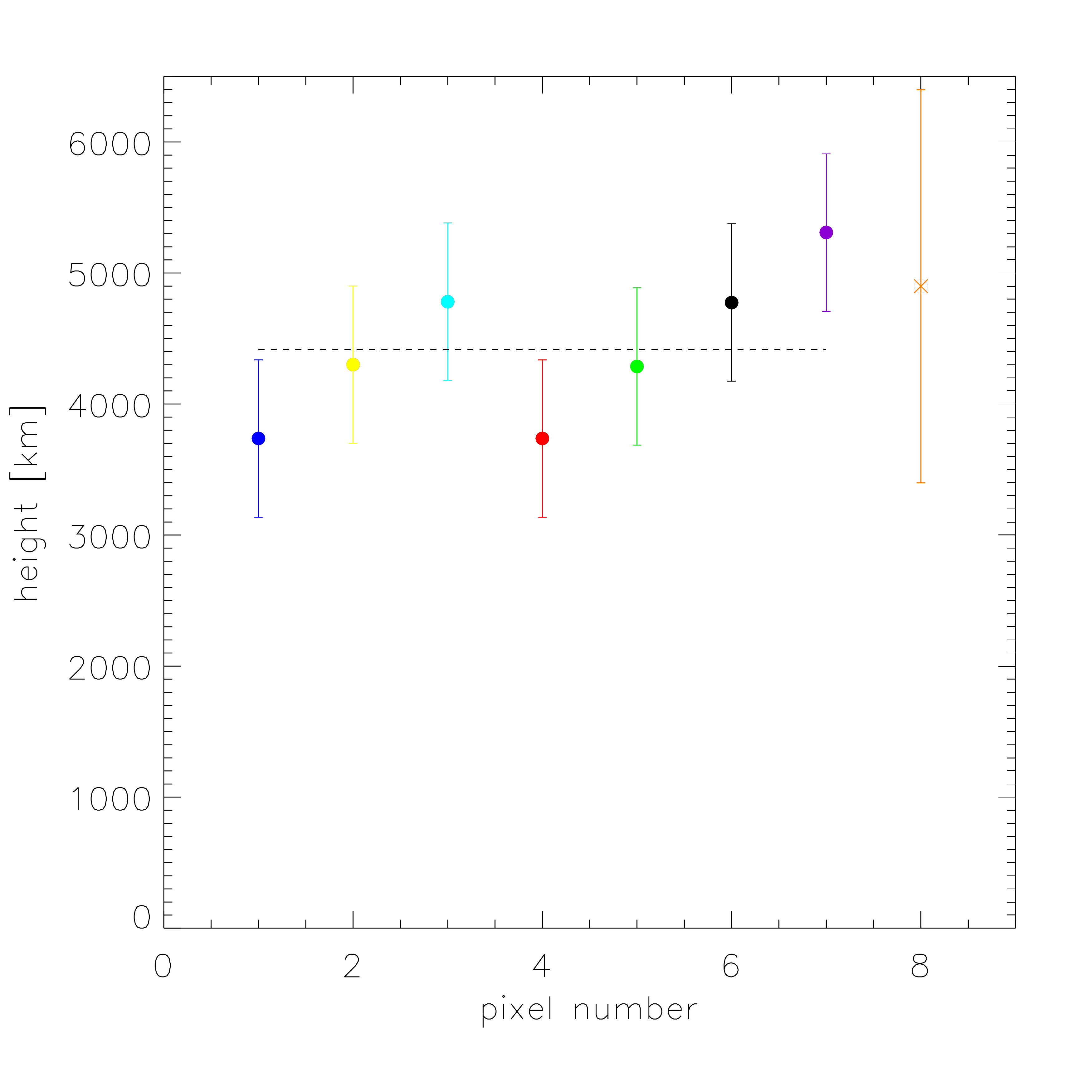}}
\end{subfigure}
\begin{subfigure}[t]{0.33\textwidth}
\centering
{ \includegraphics[width=1.0\hsize]{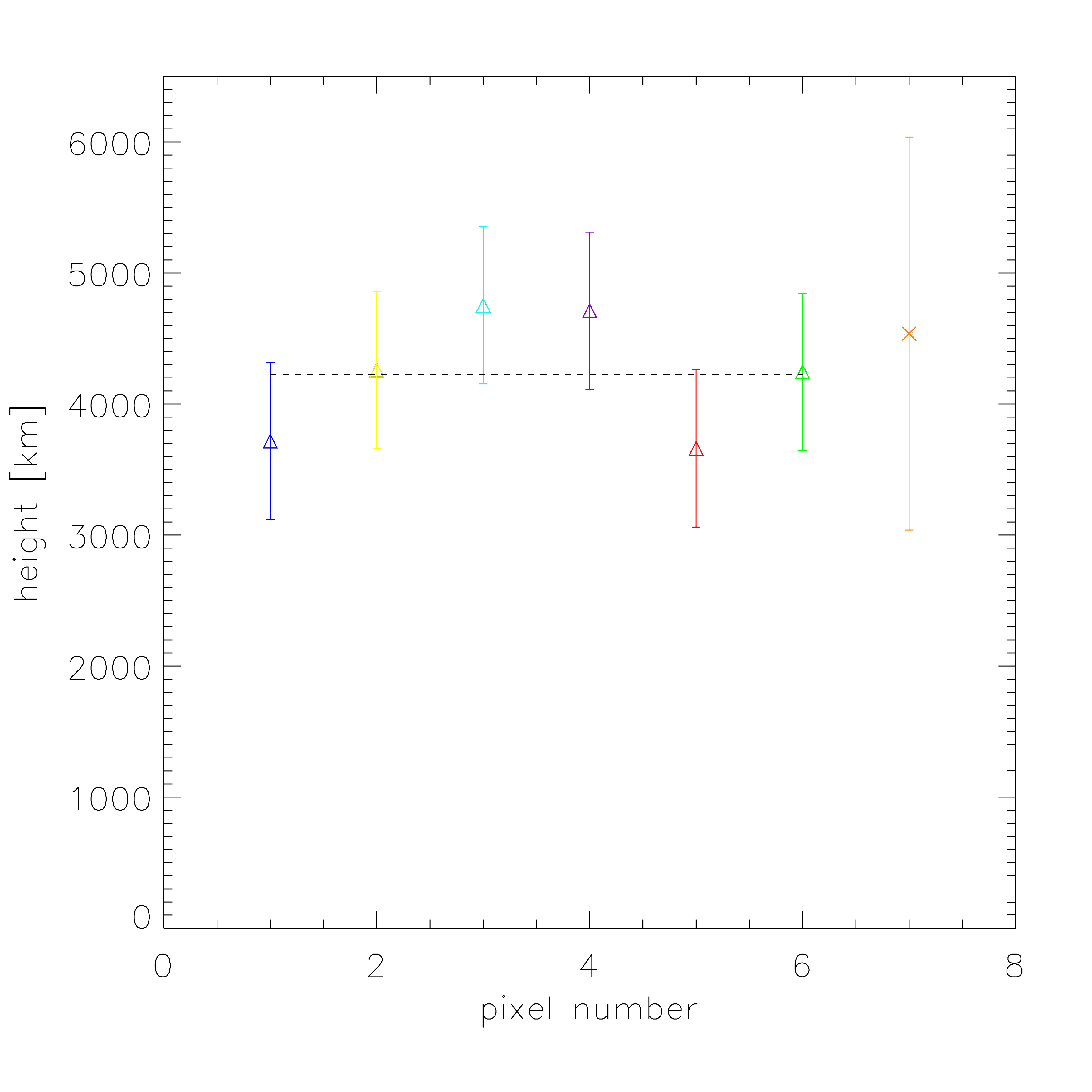}}
\end{subfigure}
\begin{subfigure}[t]{0.33\textwidth}
\centering
{ \includegraphics[width=1.0\hsize]{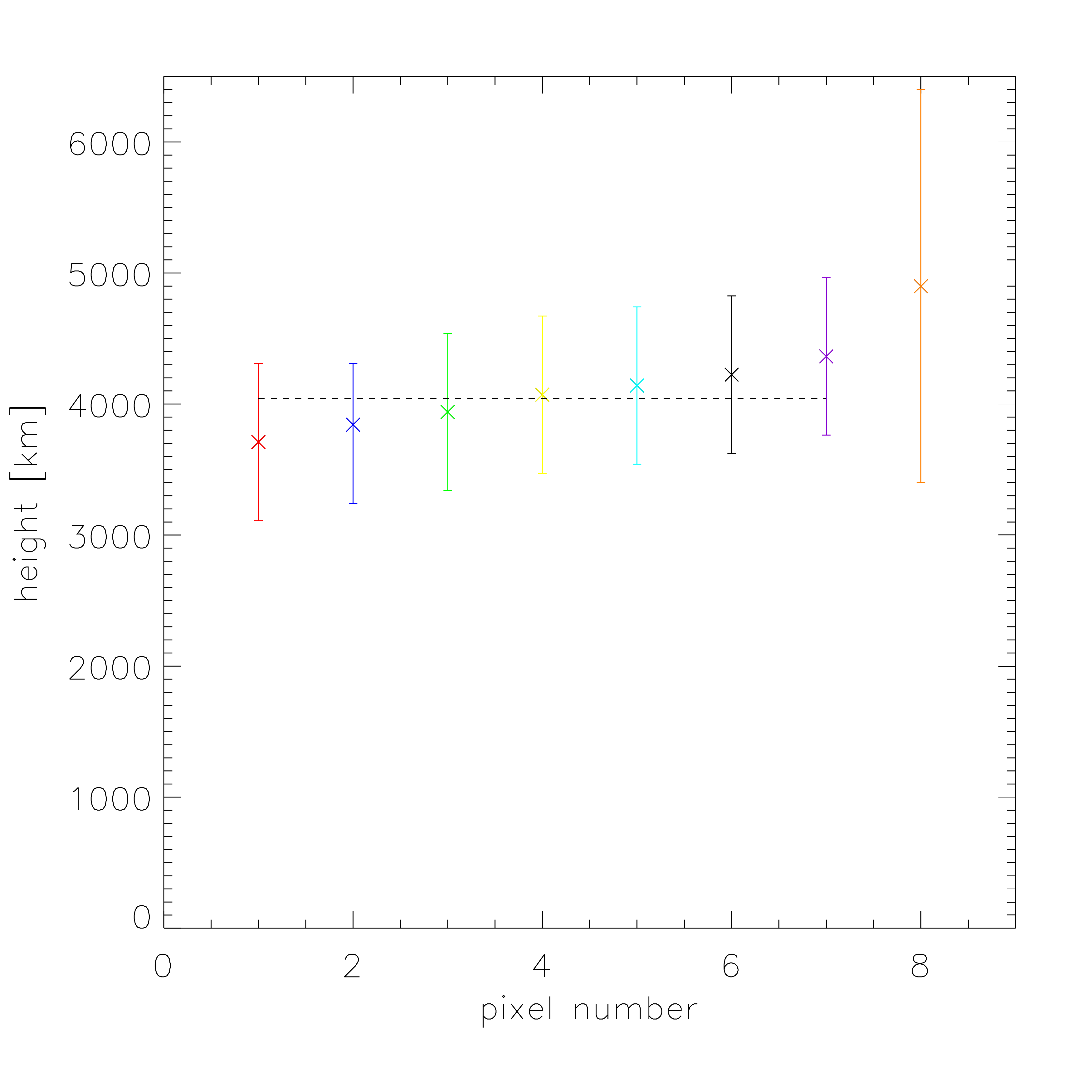}}
\end{subfigure}
\caption{Heights of individual pixels of campfire 15. Left panel: Heights of pixels marked with colored circles in the top panels of Fig.~\ref{fig:triang_hri}, together with the height of the whole campfire triangulated starting from the \hrieuv data (as reported in Fig.~\ref{fig:height_size}) shown with the orange cross. Middle panel: Heights of individual pixels marked with colored triangles in the middle panels of Fig.~\ref{fig:triang_hri}, together with the height of the whole campfire triangulated starting from the SDO/AIA data shown with the orange cross. Right panel: Heights of points marked with colored crosses along the epipolar line shown in the bottom panels of Fig.~\ref{fig:triang_hri}.  In all panels, the horizontal dashed line marks the average of individual pixels. 
}
\label{fig:heights_hri}
\end{figure*}

\begin{figure*}[ht]
\centering
{ \includegraphics[width=1.0\hsize]{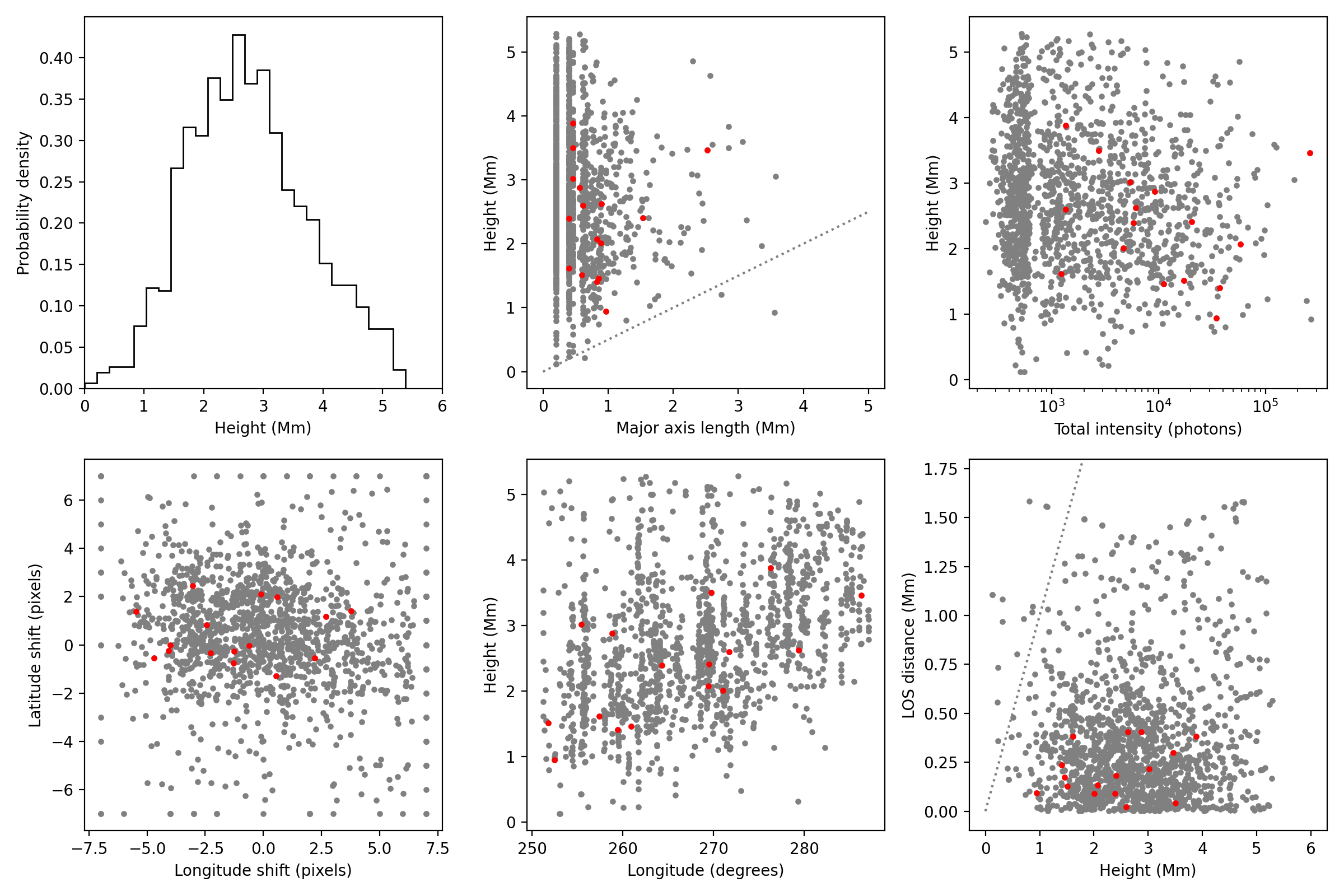}}
\caption{Results of the automatic triangulation. Top left: Histogram of triangulated campfire heights. Top middle: scatter plot of height {\it vs.}  the major axis length (the dotted line shows the $H = L/2$ relation, the red dots correspond to events triangulated manually in Sect.~\ref{section:manual}). Top right: Scatter plot of height {\it vs.}  the total intensity of a campfire. Bottom left: scatter plot of campfire shifts in latitude and longitude between the \hrieuv and AIA 171~\AA\ images. Bottom middle: Scatter plot of campfire height vs. the longitude. Bottom right: Scatter plot of the length of the shortest segment between the \hrieuv and AIA 171~\AA\ lines of sight for each campfire vs. its height. Most events lie below the dashed line indicating equal segment length and height. 
}
\label{fig:histograms}
\end{figure*}

Another error is linked to the uncertainty of the pointing of the \hrieuv telescope. The position of the center of the Sun in the image can be derived from the FITS file header keywords CRVAL1 and CRVAL2 \citep{Thompson2006}. However, those turn out to have insufficient  precision for the small-scale stereoscopy reported in this study. We carried out a parameter search to identify the best CRVAL1 and CRVAL2 values. For one of the very small campfires, we selected the brightest pixel in the \hrieuv image (Fig.~\ref{fig:manual}(a)) and verified whether the corresponding epipolar line passes through the brightest campfire pixel in the AIA image. For the default CRVAL values reported in the FITS header this is not the case, so an adjustment of CRVAL1 and CRVAL2 is needed. Even after the adjustment, there is a small range of CRVAL1 values that lead to the epipolar line passing through the corresponding AIA pixel, see panels (b), (c), and (d) of Fig.~\ref{fig:manual}. The appropriate CRVAL values should be the same for all the campfires, so we repeated the procedure for other campfires. In comparison with the CRVAL values in the FITS header, the best CRVAL2 value is offset by \SI{-16.7}{\arcsecond}, and the best CRVAL1 offset values range from \SI{-1.2}{\arcsecond} to \SI{-4.2}{\arcsecond}. For other CRVAL values, at least for one campfire, the epipolar line does not pass through the required pixel, or subphotospheric heights are found, so these values are discarded. We then perform the triangulation for four CRVAL1 values: \SI{-1.2}{\arcsecond}, \SI{-2.2}{\arcsecond}, \SI{-3.2}{\arcsecond}, and \SI{-4.2}{\arcsecond}. The two extreme offset values (\SI{-1.2}{\arcsecond} and \SI{-4.2}{\arcsecond} (see panels (b) and (d) in Fig.~\ref{fig:manual}) lead to the epipolar line passing close to the lower and upper edges of the pixel, respectively. The resulting heights are shown in the four panels of Fig.~\ref{fig:height_size}. We note that for a number of campfires, the height values for the CRVAL1 offset of \SI{-1.2}{\arcsecond} are within 100~km from the photospheric surface, which we consider unlikely. We adopt as the most likely values the heights measured for the CRVAL2 offset of \SI{-3.2}{\arcsecond}. We note that the CRVAL1 and CRVAL2 values determined in this way are very close to those obtained using the automatic method (see Sect.~\ref{section:automatic}). The shift of the CRVAL1 value by $\pm$\SI{1}{\arcsecond} leads to the shift of the campfire height by $\pm 900$~km. We adopt this value as a systematic error that has to be added on top of the random error mentioned above. 

\subsection{Results of manual triangulation}
\label{section:results_manual}

We plot the height above the photosphere versus the projected length of each campfire in Fig.~\ref{fig:height_size}. The projected length of a campfire in the image plane is defined as the maximal distance between the two farthest campfire pixels. Figure~\ref{fig:height_size} shows that campfires are situated between 1000~km and 5000~km from the photosphere\footnote{The heights reported in Fig.~6 of \citet{Berghmans2021} are slightly different from the heights in the bottom left panel of Fig.~\ref{fig:height_size}, as we now remade the triangulation using the correct position of SDO instead of assuming it is at the Earth.}. We compared the obtained dependence between the height and the length to that expected for semi-circular loops. For a semi-circular loop of  length $L$ (measured as the distance between the footpoints), the apex is situated at  height $H = L/2$. This dependence is plotted in Fig.~\ref{fig:height_size} with the black line. This is the maximal height of any point of a semi-circular loop of length $L$. Figure~\ref{fig:height_size} demonstrates that the campfires are situated systematically higher than the apex of the semi-circular loop with the same projected length. This means that the campfire loops may be elongated, that is they are taller than semi-circular loops. However, we do not expect the emission of coronal plasma at 174~\AA\ to be visible at photospheric heights, so the photospheric footpoints of campfire loops should not be visible in the 174~\AA\ passband of the \hrieuv telescope. Therefore, a more likely interpretation is that the emitting plasma of campfires is located around the apexes of corresponding loops; for more, see Fig.~7 in \citet{Berghmans2021}. 

\subsection{Dependence on the selection of points}
\label{section:resolved}

The manual method depends on the visual identification of the same pixel in two views. We now address the issue about how reliable such identification is. In order to do this, we selected a large campfire (campfire 15) that has a clear fine structure consisting of a few pixels (see Fig.~\ref{fig:triang_hri}). The orange cross indicates the position used for the triangulation of the whole campfire (Fig.~\ref{fig:height_size}). We attempt to triangulate a few pixels constituting the campfire, first starting from the \hrieuv image as it has higher resolution.

We selected the centers of seven brightest pixels of the campfire and for each of them determined the counterpart bright point along the corresponding epipolar line in the SDO/AIA 171~\AA\ image. As the campfire is aligned along the epipolar line, the process is poorly constrained, so the identification is only approximate. The result is shown in top panels of Fig.~\ref{fig:triang_hri}. The process obviously does not preserve the campfire shape, probably due to the integration along different lines of sight. Nevertheless, all seven \hrieuv campfire pixels have a counterpart bright pixel in the AIA image. The left panel of Fig.~\ref{fig:heights_hri} shows that the resulting heights are very similar, and the average height of individual pixels is very close to the height of the whole campfire triangulated as a whole starting from the \hrieuv data (as reported in Fig.~\ref{fig:height_size}). Within the errorbars, it can be considered that the campfire lies in a limited range of heights.  

We repeated the same process but starting with the AIA image. The result is shown in the middle panels of Fig.~\ref{fig:triang_hri}. It can be seen that for six campfire AIA pixels, a plausible counterpart point can be found in the bright part of the \hrieuv campfire. One point (marked with the black cross in the middle right panel of Fig.~\ref{fig:triang_hri}) cannot be mapped to a bright pixel of the \hrieuv campfire. Overall, the identification of the pixels gives a result similar to that of the process starting from the \hrieuv data, although sometimes differences up to two \hrieuv pixels could be seen. For example, positions of the two magenta pixels in the \hrieuv image in top left and middle left panels of Fig.~\ref{fig:triang_hri}, may correspond to a single AIA pixel. 

The middle panel of Fig.~\ref{fig:heights_hri} shows that the resulting heights of individual pixels are again very similar. The average height of individual pixels is somewhat lower than the height of the whole campfire triangulated as a whole starting from the AIA data, although it is with the error bar of the latter value. 

Finally, we triangulated a few pixels along the epipolar line, see bottom panels of Fig.~\ref{fig:triang_hri}. In this case, the two endpoints are determined with confidence based on the campfire morphology in the \hrieuv and AIA images. The rest of the points are placed equidistantly along the epipolar line. The heights of the points are predictably gradually increasing from the red to the magenta point. 

The general agreement (within the errorbars) between all the individual and average heights reported in Fig.~\ref{fig:heights_hri} demonstrates that, despite the poorly constrained process of selecting the triangulated points, the manual identification of pixels in \hrieuv and AIA images does not influence the triangulation process much, and the derived heights are robust.

\section{Automatic triangulation}
\label{section:automatic}

\subsection{Method and results}
\label{section:method_automatic}

Manual triangulation can become impractical if the number of events is large. In order to triangulate the height of the 1468 events reported by \citet{Berghmans2021}, we devised an automated scheme. This was achieved by measuring the parallax shift between the respective locations of campfires in \hrieuv and AIA Carrington-projected images. At the time of the observations, during commissioning, the attitude of the Solar Orbiter spacecraft was still subject to unwanted perturbations. In order to estimate the position of the center of the Sun in the images with sufficient accuracy, we minimized the squared residuals between the \hrieuv and AIA images projected to Carrington coordinates. We thus determined corrections to the CRVAL1 and CRVAL2 FITS keywords of \SI{-2.9}{\arcsecond} and \SI{-16.8}{\arcsecond} respectively. These values are very close to the values determined manually in Sect.~\ref{section:manual}. A correction to the CROTA keyword (corresponding to the roll angle of the solar rotational axis in the image) of \SI{0.045}{\arcsecond} was found too. The Carrington remapping assumes that the observed emission comes from a solid sphere. Structures lying below or above the transformation radius appear to have two different sets of Carrington coordinates as seen from the two vantage points. For each event detected in the projected \hrieuv images, we used local cross-correlation in a $29\times 29$ pixels box to determine its Carrington coordinates in the projected AIA image corresponding to the time of its maximum of intensity. The bottom left panel of Fig.~\ref{fig:histograms} is a scatter plot of the resulting \hrieuv to AIA shifts in longitude and latitude. The distribution is centered on (0, 0) because the radius of the projection sphere was chosen (a posteriori) to be $1.004~R_\odot$ (\SI{2.8}{\mega\metre} above the surface), which corresponds to the mean altitude of the detected events. We verified that changing the projection radius does not change the triangulation method described hereafter.

\begin{figure}[ht]
\centering
{ \includegraphics[width=1.0\hsize]{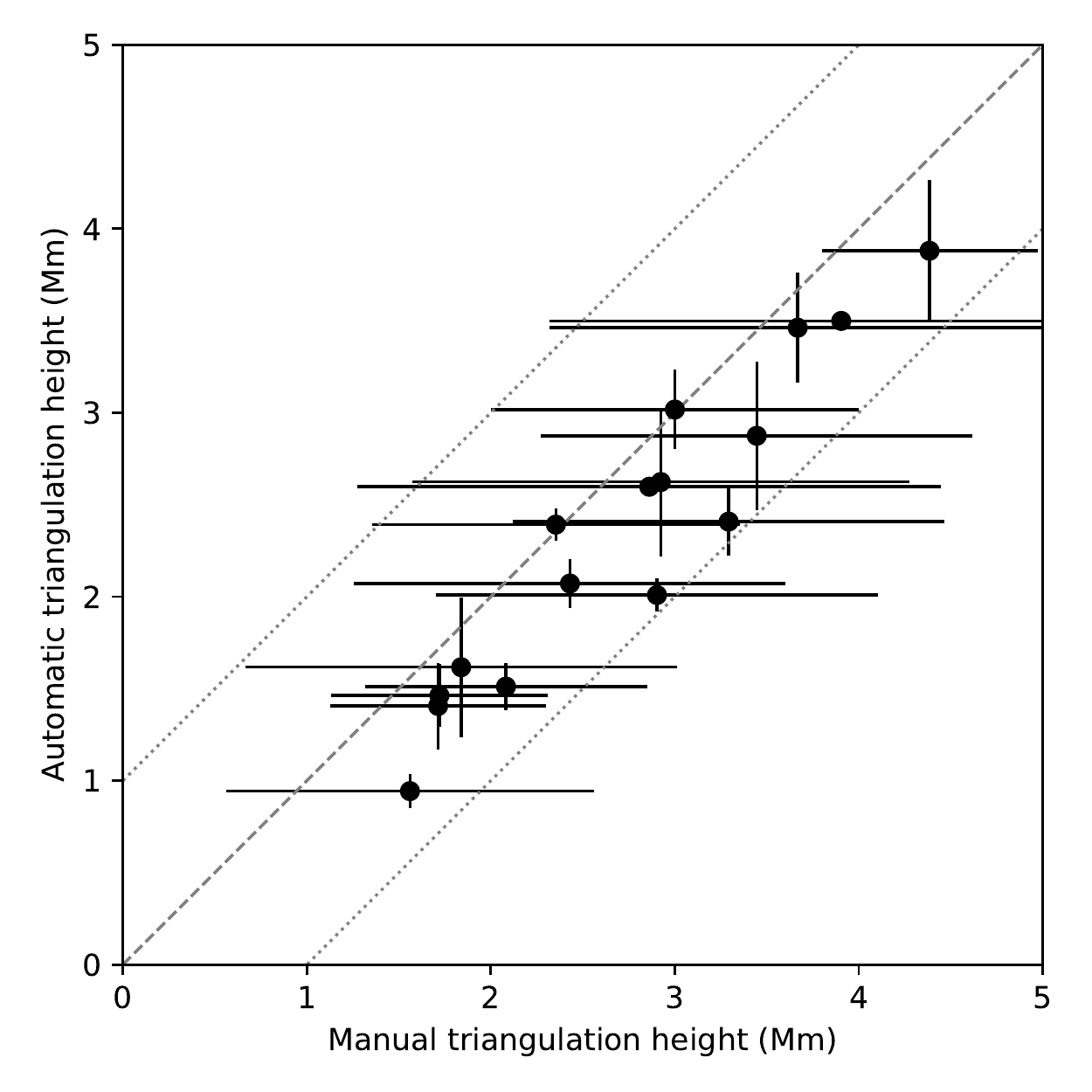}}
\caption{Comparison of the heights obtained by manual (abscissa) and  automatic (ordinate) triangulation. The two methods give remarkably consistent results. 
}
\label{fig:man-auto}
\end{figure}

For each event, the two spacecraft and the two corresponding sets of apparent coordinates on the Carrington sphere define two LOS. Since the points of maximum correlation in the AIA images are not forced to lie on the epipolar plane defined by the two spacecraft and the true 3D event location, these LOS do not necessarily intersect. They, however, generally come very close and the reported height is that of the middle point of the shortest segment joining them (Fig.~\ref{fig:histograms}, top left panel). The bottom right panel of Fig.~\ref{fig:histograms} show that the shortest distance between the two LOS is almost always smaller than the height of the middle point, which indicates that the latter is a good approximation for the height of the campfire. As a byproduct, this distance gives an indication of the reliability of the pairing of features between HRI and AIA images. The other panels of Fig.~\ref{fig:histograms} that show the height versus the longitude, major axis length and intensity do not exhibit any obvious correlation. We verified that using different cross-correlation algorithms or different widths of the correlation window do not alter the results in a significant manner.  

\subsection{Comparison of heights derived using manual and automatic methods}
\label{section:comparison}

Figure~\ref{fig:man-auto} shows the automatically triangulated heights for each of the event re-triangulated manually with the method described in Sect.~\ref{section:manual} for the same times as the automatic detection. The CRVAL1 and CRVAL2 offsets of \SI{-2.9}{\arcsecond} and \SI{-16.8}{\arcsecond} were adopted. The vertical error bars represent the length of the shortest LOS segment. The horizontal error bars are calculated as in Sect.~\ref{section:manual}. The dashed lines represent $\pm$\SI{1}{\mega\metre} around the diagonal. The points are distributed along the diagonal with a correlation coefficient of 0.86, which indicates an excellent agreement between the two methods and validates the automatic approach.

\section{Summary and discussion}
\label{section:Discussion}

In a first commissioning quiet Sun data set taken by \hrieuv, numerous small-scale EUV brightenings (dubbed ``campfires'') were discovered \citep{Berghmans2021}. At that time, the angular separation of the Solar Orbiter and SDO spacecraft was \SI{31.514}{\degree}, which allowed for the first stereoscopy of structures in the solar atmosphere at such a high resolution, namely 400~km for \hrieuv and 880~km for AIA. Using manual and automatic triangulation techniques, we determined that the campfire heights are in the range between \SI{1000}{\kilo\metre} and \SI{5000}{\kilo\metre} above the photospheric surface. The random errors are between 600~km and 1500~km, and the systematic error is around 900~km. 

For a resolved campfire consisting of a few pixels in both \hrieuv and AIA data (projected length around 3000~km, height around 4000~km), the triangulated heights of individual pixels lie in a rather narrow range of less than 2000~km. Starting the triangulation process from \hrieuv or AIA data does not influence the heights in a significant way. 

The campfire temperatures were estimated through the DEM analysis, with the DEM peaking around $\log T = 6.1$ \citep{Berghmans2021}. These are coronal temperatures, however, the campfire heights are remarkably low. The EUI observations and our triangulation provide the first evidence that coronal structures emitting around 174~\AA\ do exist at such low heights in quiet Sun regions. We compared the heights of campfires with their projected lengths and determined that the semi-circular loop geometries cannot describe the results of the triangulation: campfires are located systematically higher. This could be interpreted in two ways. First, campfires may be loops elongated in the vertical direction; however, this implies that we see the full loop length emitting at coronal temperatures, which we consider unlikely. Second, campfires may be the bright apexes of low-lying small-scale network loops that are heated to temperatures around 1~MK. This explanation is more plausible and is confirmed by a limited range of heights determined in a resolved large campfire (see Sect.~\ref{section:resolved}).

The classical description of the transition region proposed by \citet{Gabriel1976} interprets it as a thermal interface between the coronal and chromospheric plasmas. This description has been challenged by observations of a too high radiance of transition region lines. The idea of the ``unresolved fine structure'' (UFS) of the transition region resolved this problem but postulated the existence of numerous small-scale loops in the quiet Sun that have no magnetic interface to the corona \citep[see ][]{Feldman1983, Dowdy1986}. These structures would have the transition region temperatures (between $10^4$ and $10^6$~K). Following the improvement of the spatial resolution of observations, such structures were indeed detected \citep[e.g.,][]{Feldman1999, Warren2000, Hansteen2014, Chitta2021}. In particular, \citet{Hansteen2014}, using the data from the IRIS mission \citep[][spatial resolution of \SI{290}{\kilo\metre}, or \SI{0.4}{\arcsecond} at \SI{1}{\astronomicalunit}]{IRIS}, identified numerous small-scale low transition region loops constituting the UFS. Their apparent lengths were between 4 and 8~Mm, and the heights were between 1 and 5~Mm. The projected lengths and heights of \hrieuv campfires observed at hot coronal temperatures are remarkably similar to the lengths and heights of cold low transition region loops found by \citet{Hansteen2014}. This implies that hot and cold small-scale, low-lying network loops coexist in the solar transition region, confirming the theoretical concepts of \citet{Feldman1983} and \citet{Dowdy1986}. This fine structure may also be very dynamic \citep[e.g.,][]{Schmit2016}, which is confirmed by the dynamic character of some campfires, for instance interacting loops in some of them as reported by \citet{Berghmans2021}.

Coronal bright points have similar morphologies but bigger sizes than EUI campfires. Their underlying photospheric magnetic field is bipolar \citep[e.g., ][]{Madjarska2019}. As mentioned in Sect.~\ref{section:Observations}, some of the campfires are projected on photospheric bipoles, such as regular coronal bright points observed at larger scales. So their physics could be similar to the physics of coronal bright points. However, a simple bipolar field hypothesis is not applicable to all the campfires. There are several possible scenarios to explain the origin of the campfires which are not projected on photospheric bipoles. The underlying small-scale magnetic bipoles may be unresolved (spatially or temporarily). Complex magnetic field topologies with elongated field lines may be envisaged, so that an essentially bipolar coronal structure is not projected on the corresponding photospheric bipole. Another possibility is that campfires may be not related to bipoles but result from small-scale transient heating events in the large-scale magnetic field. These heating events may be produced for example by the component reconnection as proposed by \citet{Chen2021}. The resulting simulated transient brightenings have lengths of 1--4~Mm and heights of 2--5~Mm \citep{Chen2021}, which are in a good correspondence with heights and lengths of campfires reported above (see Figs.~\ref{fig:height_size} and \ref{fig:histograms}). Further detailed analysis is necessary to understand the relationship between the photospheric magnetic field and campfires.

Coronal bright points are estimated to be situated at heights from 8000~km to 12000~km \citep{Brajsa2004}. Although smaller heights down to 5000~km are sometimes reported for bright points \citep{Kwon2012, Sudar2016}, campfires are generally situated lower than coronal bright points \citep[cf. the height distributions in our Fig.~\ref{fig:histograms} and in Fig.~3 of ][]{Kwon2010}. Low-lying structures at coronal temperatures have been observed in active regions too. In particular, we recall observations of the transition region moss that has temperatures of 0.6--1.6~MK \citep{Fletcher1999} and is situated at footpoints of hot (3--5~MK) loops, at heights between 2000~km and 5000~km \citep{Berger1999, Martens2000}.  Campfires are observed in the quiet Sun, and they present morphologies that are very different from those of the moss. The moss elements do not appear as small-scale loops, and they are interspersed by cold spicules \citep{DePontieu1999}. Moss is interpreted as the classical transition region \citep{Martens2000}, but for campfires we favor the interpretation in terms of small-scale isolated loops. We note that Bifrost simulations exhibit the 1~MK emission at heights around 2000~km \citep{Martinez2017}, but they essentially show the classical transition region emission. Other simulations demonstrate that the moss emission may result from interaction of photospheric motions with the magnetic field, which is inherently dynamic and cannot be limited to the emission of plasma confined in low-lying loops \citep{Testa2016}. It is uncertain if the simulations showing the low-lying moss emission can be applicable to the quiet Sun situation. Furthermore, it is unclear why the low-lying structures at coronal temperatures are so different in active regions (moss, probably classical transition region) and in the quiet Sun (campfires, small-scale loops). This can be addressed using future Solar Orbiter observations (EUV images, photospheric magnetograms, and spectroscopic measurements) combined with state-of-the-art modeling.

The automatic method of campfire identification and triangulation opens up possibilities for statistical studies. The very similar heights found by the manual triangulation at the beginning of the sequence (14:54:10~UT) and by the automatic triangulation at the time of the maximum intensity indicate that the heights do not change much during the 5~minutes covered by the data set. Together with the fact that limited horizontal motions are detected in most of the campfires \citep{Berghmans2021}, this indicates that the campfires are brightenings of structures that likely remain confined to a narrow range of heights. The lack of vertical and horizontal motions and the characteristics of the campfire morphologies do not support the interpretation of most of the campfires in terms of top-down viewed jets, spicule side-effects, or upward-propagating waves. It could be possible that faint jet-like structures may be rooted in some campfires, although only dedicated case studies could confirm or refute this possibility. Subsequent \hrieuv observations, in particular, those made from the Solar Orbiter perihelia situated as close as \SI{0.29}{\astronomicalunit} will provide more information about the structure of the transition region and the corona, and further clarify the role of campfires in the physical processes in the solar atmosphere. 


\begin{acknowledgements}
A.~N.~Z. dedicates this work to the memory of his teacher and mentor, Igor S. Veselovsky (1940--2020), who was an ardent proponent of studies of the coronal fine structure. Solar Orbiter is a space mission of international collaboration between ESA and NASA, operated by ESA. The EUI instrument was built by CSL, IAS, MPS, MSSL/UCL, PMOD/WRC, ROB, LCF/IO with funding from the Belgian Federal Science Policy Office (BELPSO); the Centre National d’Etudes Spatiales (CNES); the UK Space Agency (UKSA); the Bundesministerium für Wirtschaft und Energie (BMWi) through the Deutsches Zentrum für Luft- und Raumfahrt (DLR); and the Swiss Space Office (SSO). The ROB team thanks the Belgian Federal Science Policy Office (BELSPO) for the provision of financial support in the framework of the PRODEX Programme of the European Space Agency (ESA) under contract numbers 4000112292, 4000117262, and 4000134474. PA and DML acknowledge funding from STFC Ernest Rutherford Fellowships No. ST/R004285/2 and ST/R003246/1, respectively. SP acknowledges the funding by CNES through the MEDOC data and operations center. LH and KB are grateful to the SNF for the funding of the project number 200021\_188390.
\end{acknowledgements}

\bibliographystyle{aa} 


\begin{appendix}

\onecolumn

\section{The stereoscopic reconstruction}
\label{app}

For each spacecraft, the LOS associated with an image pixel
$(x,y)$ can be described as
\begin{equation}
   \vect{r}(\alpha)=\vect{p} + \alpha \hat{\vect{v}}(x,y), 
\label{app:def_LOS}\end{equation}
where $\vect{r}$ and $\vect{p}$ are the positions of the triangulated structure and the observing spacecraft
in a heliocentric coordinate system,
$\hat{\vect{v}}$ the view direction associated with pixel at image
coordinates $(x,y)$, and $\alpha\in \mathbb{R}_+$ parametrizes the LOS (the hat hereafter denotes the vector of unit length). The vector
$\hat{\vect{v}}$ combines all the information on the camera system
and the spacecraft attitude which is assumed known. In particular,
we can split the view direction into the direction to the Sun's center
$-\hat{\vect{p}}$ and a vector of $\vect{t}(x,y)$ in the image plane.
If the Sun's center is at image pixel $(0,0)$, then $\hat{\vect{p}}$
and $\vect{t}$ are perpendicular and for an aptly corrected image
\begin{equation}
  \hat{\vect{v}}(x,y) = \frac{-\hat{\vect{p}}+\vect{t}}
      {||\hat{\vect{p}}-\vect{t}||},\qquad
  \vect{t}=\frac{\Delta_\mathrm{pix}}{f}\,\mathcal{R}\rvecc{x}{y},
\label{app:def_imagevec}\end{equation}
where $\Delta_\mathrm{pix}$ is the pixel size, $f$ the camera focal length
and $\mathcal{R}$ maps an image vector into the 3D heliocentric coordinate
system of $\vect{p}$. Typically,
$||\vect{t}||$ is on the order of $\sqrt{x^2+y^2}\Delta_\mathrm{pix}/f$
and much smaller than unity. For solar observations from 1~au, it is on the
order of 1/200.

If a position, $\vect{r,}$ on the solar surface is sought, a single image is
sufficient. The condition $||\vect{r}(\alpha)||^2=R^2_\odot$ imposed on
(\ref{app:def_LOS}) in this case yields a quadratic polynomial in $\alpha$ with known
coefficients, which either gives two solutions: one for the front- and
backside of the Sun or no solution if $\hat{\vect{v}}$ is directed beyond
the solar disc.

If stereoscopy is to be applied to an object at $\vect{r}$ we need the
observation from two spacecraft (without loss of generality, we call them A and B having the STEREO mission in mind). Then from
(\ref{app:def_LOS}):
\begin{equation}
   \vect{r} \simeq \vect{p}_A + \alpha_A \hat{\vect{v}}_A(x_A,y_A) 
            \simeq \vect{p}_B + \alpha_B \hat{\vect{v}}_B(x_B,y_B). 
\label{app:stereoeq}\end{equation}
Equality here only applies if the image data $(x_A,y_A)$ and $(x_B,y_B)$ are
consistent, that is, the two LOS properly intersect.
Even if this does not hold exactly, we can use (\ref{app:stereoeq}) for
a least-square solution for the unknown distances $\alpha_A$ and $\alpha_B$
along the respective LOS. Following straightforward algebra,
\begin{gather}
   \rvecc{\alpha_A}{\alpha_B}
   =-\left(\vect{V}\tp\vect{V}\right)^{-1}\vect{V}\tp(\vect{p}_A-\vect{p}_B).
\label{app:stereosol}
\end{gather}
Here, $\vect{V}=(\hat{\vect{v}}_A \;\hat{\vect{v}}_B)$ is a $3\times2$ matrix with columns $\hat{\vect{v}}_A$ and
$\hat{\vect{v}}_B$ and the superscript $\tp$ denotes the transpose.
The distance vector $\vect{p}_A-\vect{p}_B$ between the observing
spacecraft is the stereoscopy base. If the LOS from A and B properly
intersect, the columns of $\vect{V}$ and the stereoscopy base define the epipolar
plane of the reconstruction problem. This plane mapped into each
image results in a pair of associated epipolar lines. If the image coordinates
$(x_A,y_A)$ and $(x_B,y_B)$ are selected exactly from associated epipolar lines,
then the view directions $\hat{\vect{v}}_A$ and $\hat{\vect{v}}_B$ intersect.
If they do not, for example, due to camera parameter errors
or data noise, the solution (\ref{app:stereosol}) is least-square in the sense
that it yields two points, one along the LOS from either spacecraft, which have
the closest possible mutual distance. The best guess of $\vect{r}$ then is the
arithmetic average of both solutions.

Since the columns of $\vect{V}$ are normalized, we can write
(\ref{app:stereosol}) more explicitly. Using
\begin{gather*}
  \vect{V}\tp\vect{V}=\begin{pmatrix}
                     1 & \hat{\vect{v}}_A\tp\hat{\vect{v}}_B \\
                      \hat{\vect{v}}_A\tp\hat{\vect{v}}_B & 1
                      \end{pmatrix},
\nonumber\\
  \left(\vect{V}\tp\vect{V}\right)^{-1}=
                      \frac{-1}{1-(\hat{\vect{v}}_A\tp\hat{\vect{v}}_B)^2}
                      \begin{pmatrix}
                      1 & -\hat{\vect{v}}_A\tp\hat{\vect{v}}_B \\
                      -\hat{\vect{v}}_A\tp\hat{\vect{v}}_B & 1
                      \end{pmatrix},
\end{gather*}
the final result obtained for the unknown distance along the LOS is,
for example, for spacecraft A,

\begin{equation}
  \alpha_A
   = -\hat{\vect{v}}_A\tp
       \frac{1-\hat{\vect{v}}_B\hat{\vect{v}}_B\tp}
            {1-(\hat{\vect{v}}_A\tp\hat{\vect{v}}_B)^2}
                     (\vect{p}_A-\vect{p}_B).
\label{app:alpha}\end{equation}
The result for B is identical but with A and B are interchanged.

We note that for $\vect{t}=0$ we have  $\hat{\vect{v}}_{A,B}=
-\hat{\vect{p}}_{A,B}$ and (\ref{app:alpha}) gives
\begin{gather}
    \alpha_A\hat{\vect{v}}_A(0,0)
  =-\hat{\vect{p}}_A\hat{\vect{p}}_A\tp
               \frac{1-\hat{\vect{p}}_B\hat{\vect{p}}_B\tp}
                    {1-(\hat{\vect{p}}_A\tp\hat{\vect{p}}_B)^2}
                      (\vect{p}_A-\vect{p}_B)
=-\hat{\vect{p}}_A\hat{\vect{p}}_A\tp
       \frac{1-\hat{\vect{p}}_B\hat{\vect{p}}_B\tp}
            {1-(\hat{\vect{p}}_A\tp\hat{\vect{p}}_B)^2}\vect{p}_A
 =-\vect{p}_A 
       \hat{\vect{p}}_A\tp
       \frac{1-\hat{\vect{p}}_B\hat{\vect{p}}_B\tp}
            {1-(\hat{\vect{p}}_A\tp\hat{\vect{p}}_B)^2}
            \hat{\vect{p}}_A 
 =-\vect{p}_A,
\label{app:zeroorder}\end{gather}
which then results in $\vect{r}=0$.
So, for observations from a large distance, we have
$||\vect{t}||\ll1$, $||\vect{r}||\ll||\vect{p}||$ and
when (\ref{app:alpha}) is inserted into (\ref{app:def_LOS}) we
subtract two large vectors to obtain a relatively small difference
$\vect{r}$. This causes some loss of numerical precision if
the stereoscopy base $||\vect{p}_A-\vect{p}_B||$ becomes much shorter
than $||\vect{p}||$. Formally, this is caused by a bad condition
of $\vect{V}\tp\vect{V}$ in (\ref{app:stereosol}), with its determinant
$1-(\hat{\vect{v}}_A\tp\hat{\vect{v}}_B)^2 \simeq 
 1-(\hat{\vect{p}}_A\tp\hat{\vect{p}}_B)^2=\sin^2\gamma$
decreasing to zero as $\gamma$, the heliocentric angular distance
between the two observing spacecraft, shrinks.
We found that this potential concern on precision is minor compared
with the uncertainties caused the limited precision of our data and
observational parameters.
For $||\vect{r}||\simeq 1 R_\odot$ and $||\vect{p}||\simeq 200 R_\odot$,
the numerical error in (\ref{app:stereosol}) scales with
$10^{-11}\times(3^\circ/\gamma)^2$. A big computational advantage of
(\ref{app:stereosol}) is, on the other hand, that no trigonometric
functions need to be evaluated.



For estimating the errors in $\vect{r}$ due to uncertainties in $\vect{t}$ we
linearize (\ref{app:alpha}) with respect to $\vect{t}$. It is convenient to evaluate
the derivative $d\vect{r}/d\vect{t}$ at $\vect{t}=0$. Therefore, when we talk
about the epipolar plane we also no longer make any distinction between the exact
epipolar plane defined by $\hat{\vect{v}}_A$ and $\hat{\vect{v}}_B$ and the
nearby plane, spanning between $\hat{\vect{p}}_A$ and $\hat{\vect{p}}_B$.
For the error estimation, this approximation is acceptable if
we observe from a distance $||\vect{p}||\gg||\vect{r}||$.
For the linearization, we use:
\begin{gather*}
  ||\hat{\vect{p}}_A-\vect{t}_A|| = 1+ t^2_A = 1 + \mathcal{O}(t^2),
\\
     \hat{\vect{v}}_A\tp\hat{\vect{v}}_B
  =  \hat{\vect{p}}_A\tp\hat{\vect{p}}_B
   - (\hat{\vect{p}}_A\tp\vect{t}_B+\hat{\vect{p}}_B\tp\vect{t}_A)
  + \mathcal{O}(t^2),
\\
    (\hat{\vect{v}}_A\tp\hat{\vect{v}}_B)^2
  = (\hat{\vect{p}}_A\tp\hat{\vect{p}}_B)^2
  - 2(\hat{\vect{p}}_A\tp\hat{\vect{p}}_B)
     (\hat{\vect{p}}_A\tp\vect{t}_B+\hat{\vect{p}}_B\tp\vect{t}_A)
  + \mathcal{O}(t^2),
\\
  \frac{1}{1-(\hat{\vect{v}}_A\tp\hat{\vect{v}}_B)^2}
  = \frac{1}{1-(\hat{\vect{p}}_A\tp\hat{\vect{p}}_B)^2}
   [1-\frac{2\hat{\vect{p}}_A\tp\hat{\vect{p}}_B}
            {1-(\hat{\vect{p}}_A\tp\hat{\vect{p}}_B)^2}
  (\hat{\vect{p}}_A\tp\vect{t}_B+\hat{\vect{p}}_B\tp\vect{t}_A)]
  + \mathcal{O}(t^2),
\\
    \hat{\vect{v}}_A\tp\hat{\vect{v}}_B
  = \hat{\vect{p}}_A\tp\hat{\vect{p}}_B
        +\hat{\vect{p}}_A\tp\vect{t}_B 
        +\hat{\vect{p}}_B\tp\vect{t}_A + \mathcal{O}(t^2),
 \\
 \hat{\vect{v}}_A\hat{\vect{v}}_A\tp
 =\hat{\vect{p}}_A\hat{\vect{p}}_A\tp
  + \vect{t}_A\hat{\vect{p}}_A\tp 
  + \hat{\vect{p}}_A\vect{t}_A\tp+ \mathcal{O}(t^2),
\end{gather*}
where $1-(\hat{\vect{p}}_A\tp\hat{\vect{p}}_B)^2=\sin^2\gamma$. Terms such as
$\hat{\vect{v}}$ or $\hat{\vect{v}}\hat{\vect{v}}$ occur at three instances
in (\ref{app:alpha}).
By the chain rule and using (\ref{app:zeroorder}), we obtain three terms:
\begin{gather*}
  \Delta\vect{t}\frac{d\vect{r}}{d\vect{t}}
   = \Delta\vect{r}=\Delta(\alpha_A\hat{\vect{v}}_A)
   =  (\Delta\vect{t}_A\hat{\vect{p}}_A\tp
      +\hat{\vect{p}}_A\Delta\vect{t}_A\tp)
      \underbrace{\frac{1-\hat{\vect{p}}_B\hat{\vect{p}}_B\tp}
                       {1-(\hat{\vect{p}}_A\tp\hat{\vect{p}}_B)^2}}_%
      {\perp \hat{\vect{p}}_B}   (\vect{p}_A-\vect{p}_B)
   -\hat{\vect{p}}_A\hat{\vect{p}}_A\tp
       \frac{\Delta\vect{t}_B\hat{\vect{p}}_B\tp
            +\hat{\vect{p}}_B\Delta\vect{t}_B\tp}
            {1-(\hat{\vect{p}}_A\tp\hat{\vect{p}}_B)^2}
                     (\vect{p}_A-\vect{p}_B)
\\          +2\hat{\vect{p}}_A(\hat{\vect{p}}_A\tp\hat{\vect{p}}_B)
      \frac{(\hat{\vect{p}}_A\tp\Delta\vect{t}_B
            +\hat{\vect{p}}_B\tp\Delta\vect{t}_A)}
   {(1-(\hat{\vect{p}}_A\tp\hat{\vect{p}}_B)^2)^2}
   + \mathcal{O}(\Delta t^2)   =   \Delta\vect{t}_A
      \underbrace{\frac{(\hat{\vect{p}}_A\tp\vect{p}_A)
                       -(\hat{\vect{p}}_A\tp\hat{\vect{p}}_B)
                        (\hat{\vect{p}}_B\tp\vect{p}_A)}
                     {1-(\hat{\vect{p}}_A\tp\hat{\vect{p}}_B)^2}}_{=p_A}
   \;+\;\xi\hat{\vect{p}}_A,
\end{gather*}
where $\xi$ is a scalar. We therefore have a single term along $\Delta\vect{t}_A$, all the other
terms are along $\vect{p}_A$ and yield the depth error.
Instead of determining the complicated expression along $\hat{\vect{p}}_A$,
we use the analogue result for spacecraft B and project both into the
respective image plane. This gives four relations, which is more than enough to
solve for $\Delta\vect{r}$ (recall that $\vect{t}_A$ is perpendicular
to $\hat{\vect{p}}_A$ and $\vect{t}_B$ is perpendicular
to $\hat{\vect{p}}_B$):
\begin{equation}
   (1-\hat{\vect{p}}_A\hat{\vect{p}}_A\tp)\Delta\vect{r}
  =p_A\Delta\vect{t}_A,
\qquad
   (1-\hat{\vect{p}}_B\hat{\vect{p}}_B\tp)\Delta\vect{r}
  =p_B\Delta\vect{t}_B.
\label{app:perperror}\end{equation}
The solution is simple if we introduce a local coordinate frame with
the three axes:
\begin{align*}
  \hat{\vect{n}}&=\frac{\hat{\vect{p}}_A\times\hat{\vect{p}}_B}
                     {||\hat{\vect{p}}_A\times\hat{\vect{p}}_B||}
               &&\simeq\text{normal to the epipolar plane,}\\
  \hat{\vect{p}}_+&=\frac{\hat{\vect{p}}_A+\hat{\vect{p}}_B}
                       {||\hat{\vect{p}}_A+\hat{\vect{p}}_B||}
               &&\simeq\text{direction half way between the two LOS,}\\
  \hat{\vect{p}}_-&=\frac{\hat{\vect{p}}_A-\hat{\vect{p}}_B}
                       {||\hat{\vect{p}}_A-\hat{\vect{p}}_B||}
               &&\simeq\text{perpendicular to $\hat{\vect{p}}_+$ in the
                        epipolar plane.}
\end{align*}
The approximate equality indicates here that the plane spanned by
$\hat{\vect{p}}_A$ and $\hat{\vect{p}}_B$ is only approximately the
exact epipolar plane as mentioned above.
The three unit directions are obviously orthogonal and their action
on the projections in
(\ref{app:perperror}) have the following properties:
\begin{gather*}
  \hat{\vect{n}}(1-\hat{\vect{p}}_A\hat{\vect{p}}_A\tp)=\hat{\vect{n}},
\qquad
  \hat{\vect{n}}(1-\hat{\vect{p}}_B\hat{\vect{p}}_B\tp)=\hat{\vect{n}},
\\  
  \hat{\vect{p}}_+[(1-\hat{\vect{p}}_A\hat{\vect{p}}_A\tp)
                  +(1-\hat{\vect{p}}_B\hat{\vect{p}}_B\tp)]
  =2\hat{\vect{p}}_+-\hat{\vect{p}}_+
                   (1+(\hat{\vect{p}}_A\tp\hat{\vect{p}}_B))
  =\hat{\vect{p}}_+(1-\hat{\vect{p}}_A\tp\hat{\vect{p}}_B),
\\  
  \hat{\vect{p}}_-[(1-\hat{\vect{p}}_A\hat{\vect{p}}_A\tp)
                  +(1-\hat{\vect{p}}_B\hat{\vect{p}}_B\tp)]
  =2\hat{\vect{p}}_--\hat{\vect{p}}_-(1-(\hat{\vect{p}}_A\tp\hat{\vect{p}}_B))
  =\hat{\vect{p}}_-(1+\hat{\vect{p}}_A\tp\hat{\vect{p}}_B).
\end{gather*}
Multiplying (\ref{app:perperror}) with $\hat{\vect{n}}$ yields two error
estimates in the direction normal to the epipolar plane. The error that is most acceptable is the smaller of the two because the better observation gives the
final limit:
\begin{equation}
  \hat{\vect{n}}\tp\Delta\vect{r}
  =\mathrm{min}(p_A\hat{\vect{n}}\tp\Delta\vect{t}_A,
                p_B\hat{\vect{n}}\tp\Delta\vect{t}_B).
\end{equation}
The error in the epipolar plane is obtained by multiplication
of (\ref{app:perperror}) with $\hat{\vect{p}}_+$ and adding the
expressions for A and B and likewise with $\hat{\vect{p}}_-$. This yields:
\begin{equation}
  \hat{\vect{p}}_+\tp\Delta\vect{r}
 =\hat{\vect{p}}_+\tp\;\frac{p_A\Delta\vect{t}_A+p_B\Delta\vect{t}_B}
                          {1-\hat{\vect{p}}_A\tp\hat{\vect{p}}_B},
\qquad
  \hat{\vect{p}}_-\tp\Delta\vect{r}
 =\hat{\vect{p}}_-\tp\;\frac{p_A\Delta\vect{t}_A+p_B\Delta\vect{t}_B}
                          {1+\hat{\vect{p}}_A\tp\hat{\vect{p}}_B}.
\end{equation}
The first term, $\hat{\vect{p}}_+\tp\Delta\vect{r}$ is the error
in depth which scales with 
$1/(1-\hat{\vect{p}}_A\tp\hat{\vect{p}}_B)
=1/(1-\cos\gamma)=\sin^{-2}(\gamma/2)/2$.
It is larger than the error
$\hat{\vect{p}}_-\tp\Delta\vect{r}$ in the
perpendicular direction, which scales with 
$1/(1+\hat{\vect{p}}_A\tp\hat{\vect{p}}_B)
=1/(1+\cos\gamma)=\cos^{-2}(\gamma/2)/2$.

Here, we only require the error in altitude which is the length of the
position error $\Delta\vect{r}$ projected on the local radial direction of
$\vect{r}$. In Eq.~(\ref{equation:error}) of the main text, we use -- as a slightly simplified
measure for the altitude error -- the length of
$\Delta\vect{r}$ projected into the epipolar plane.
It is derived by projecting both sides of (\ref{app:perperror})
into the epipolar plane. For that purpose, we replace $\Delta\vect{r}$
and $\Delta\vect{t}_{A,B}$ by their respective projections:
\begin{equation*}
  \Delta\vect{h}
  =(1-\vect{n}\vect{n}\tp)\Delta\vect{r},\quad
  \Delta\vect{s}_A
  =(1-\vect{n}\vect{n}\tp)p_A\Delta\vect{t}_A,\quad
  \Delta\vect{s}_B
  =(1-\vect{n}\vect{n}\tp)p_B\Delta\vect{t}_B
\end{equation*}
in (\ref{app:perperror}).
We note that, for example, $\Delta\vect{s}_A$ is the measurement error along the
epipolar line in image A but magnified by $p_A$, that is, projected into the POS
of image A. We make the ansatz
$\Delta\vect{h}=\rho_A\hat{\vect{p}}_A+\rho_B\hat{\vect{p}}_B$
and solve for the unknown coefficients $\rho_{A,B}$
by multiplying both sides of the projected (\ref{app:perperror}) with
$\hat{\vect{p}}_{B}\tp$ and $\hat{\vect{p}}_{A}\tp$, respectively.
As an example, for A, we have:
\begin{gather*}
   \hat{\vect{p}}_A\tp(1-\hat{\vect{p}}_B\hat{\vect{p}}_B\tp)\Delta\vect{h}
  =\hat{\vect{p}}_A\tp(1-\hat{\vect{p}}_B\hat{\vect{p}}_B\tp)
  (\rho_B\hat{\vect{p}}_B+\rho_A\hat{\vect{p}}_A)
  =\rho_A(1-(\hat{\vect{p}}_B\tp\hat{\vect{p}}_A)^2)
  =\hat{\vect{p}}_A\tp\Delta\vect{s}_B
  \\\text{or}\quad
  \rho_A=\frac{\hat{\vect{p}}_A\tp\Delta\vect{s}_B}
              {1-(\hat{\vect{p}}_B\tp\hat{\vect{p}}_A)^2}
  =\frac{\hat{\vect{p}}_A\tp\Delta\vect{s}_B}{\sin^2\gamma}.
\end{gather*}
Since the error vector $\Delta\vect{s}_B$ lies in the epipolar plane and
is perpendicular to $\hat{\vect{p}}_B$ we have, disregarding the sign
of the error vector, $\hat{\vect{p}}_A\tp\Delta\vect{s}_B
=|\Delta\vect{s}_B|\sin\gamma$.
The result for $\rho_B$ is the same with A and B interchanged.
For the squared length of $\Delta\vect{h}$ we then find
\begin{gather}
  (\Delta h)^2
   = \rho_A^2+\rho_B^2
   +2\rho_A\rho_B(\hat{\vect{p}}_A\tp\hat{\vect{p}}_B)
     = \frac{ (\hat{\vect{p}}_B\tp\Delta\vect{s}_A)^2
             +(\hat{\vect{p}}_A\tp\Delta\vect{s}_B)^2
              +2(\hat{\vect{p}}_B\tp\Delta\vect{s}_A)
                (\hat{\vect{p}}_A\tp\Delta\vect{s}_B)
              (\hat{\vect{p}}_A\tp\hat{\vect{p}}_B)}
           {\sin^4\gamma}
     = \frac{ (\Delta\vect{s}_A)^2
             +(\Delta\vect{s}_B)^2
              +2\Delta\vect{s}_A\Delta\vect{s}_B\cos\gamma}
           {\sin^2\gamma}.
\label{app:heighterr}\end{gather}

In Eq.~(\ref{equation:error}), we show half of this full error ($\delta h = \Delta h/2$).

\end{appendix}

\end{document}